\documentclass[preprint,12pt,authoryear]{elsarticle}

\makeatletter
\def\ps@pprintTitle{%
  \let\@oddhead\@empty
  \let\@evenhead\@empty
  \let\@oddfoot\@empty
  \let\@evenfoot\@oddfoot
}
\makeatother

\usepackage{amssymb,mathtools,xcolor,hyperref,cleveref,tcolorbox,adjustbox}
\graphicspath{{./figures/}}
\usepackage[ruled,linesnumbered]{algorithm2e}

\begin{document}

\begin{frontmatter}

\title{Probabilistic digital twins for geotechnical design and construction}

\affiliation[gni_label]{organization={Georg Nemetschek Institute
Artificial Intelligence for the Built World, Technical University of Munich},
            addressline={Walther-von-Dyck-Straße 10},
            city={Garching bei München},
            postcode={85748},
            country={Germany}}
\affiliation[era_label]{organization={Engineering Risk Analysis Group \& Munich Data Science Institute, Technical University of Munich},
            addressline={Theresienstr. 90},
            city={Munich},
            postcode={80333},
            country={Germany}}

\author[gni_label]{Dafydd Cotoarbă\corref{cor}}
\ead{dafydd.cotoarba@tum.de}
\author[era_label]{Daniel Straub}
\author[gni_label]{Ian FC Smith}

\cortext[cor]{Corresponding author}

\begin{abstract}

The digital twin approach has gained recognition as a promising solution to the challenges faced by the Architecture, Engineering, Construction, Operations, and Management (AECOM) industries. However, its broader application across AECOM sectors remains limited. One significant obstacle is that traditional digital twins rely on deterministic models, which require deterministic input parameters. This limits their accuracy, as they do not account for the substantial uncertainties inherent in AECOM projects. These uncertainties are particularly pronounced in geotechnical design and construction. To address this challenge, we propose a Probabilistic Digital Twin (PDT) framework that extends traditional digital twin methodologies by incorporating uncertainties, and is tailored to the requirements of geotechnical design and construction. The PDT framework provides a structured approach to integrating all sources of uncertainty, including aleatoric, data, model, and prediction uncertainties, and propagates them throughout the entire modeling process. To ensure that site-specific conditions are accurately reflected as additional information is obtained, the PDT leverages Bayesian methods for model updating. The effectiveness of the probabilistic digital twin framework is showcased through an application to a highway foundation construction project, demonstrating its potential to improve decision-making and project outcomes in the face of significant uncertainties.
\end{abstract}


\end{frontmatter}


\section{Introduction}
\label{sec:introduction}

The digital twin (DT) concept, introduced by \citep{grieves2002plm}, emerged in response to a changing reality, where the amount of data collected across all industries vastly increased due to technological advancements. Digital methods are required for real-time data processing and decision-making to leverage this for productivity gains. Thus, the digital twin shows potential for addressing challenges confronting the Architecture, Engineering, Construction, Operations, and Management (AECOM) industries. These challenges include managing the increasing flow of project-related data, low productivity, unpredictability in terms of costs and schedules, and complexity attributed to structural fragmentation \citep{opoku2021digital}.

While the digital twin concept has been adopted to address these challenges in some areas of AECOM, e.g., for facility management with the control of heating, ventilation, and air conditioning \citep{xie2023digital}, its broader application remains limited. One significant obstacle is that projects in these industries are characterized by significant uncertainties resulting from the unique and complex nature of projects. These uncertainties are particularly large in the geotechnical phase of construction projects due to the inherent variability of soil and the limited availability of observational data \citep{phoon2022geotechnical}.

Traditional digital twin approaches rely on deterministic input parameters and models to predict the behavior of the physical twin. The accuracy of the traditional digital twin depends on having complete knowledge of the physical asset, which is unattainable most of the time. Uncertainties arising from a lack of knowledge cannot be incorporated. Instead, uncertainties are reduced to deterministic values (e.g., assuming worst-case scenarios to be representative of the entire system), resulting in a loss of valuable information. This can significantly reduce the accuracy of the prediction, leaving the digital twin with no clear advantage over traditional modeling methods. Additionally, in practice, safety factors are applied to address remaining uncertainties. This might lead to overconservative designs and the inefficient usage of resources.

In recent years, there have been proposals to extend traditional digital twins with uncertainties that allow the integration Bayesian methods, and leverage their capabilities for integration additional data. Knowledge from previous projects or domain expertise is incorporated as prior belief states, which can be updated as as additional information is obtained. Previous approaches can be found under various names such as Predictive Digital Twin \citep{kapteyn2021probabilistic,chaudhuri2023predictive,torzoni2024digital}, Probabilistic Digital Twin \citep{nath2022probabilistic,agrell2023optimal} or Digital Twin Concepts with Uncertainty \citep{kochunas2021digital}. However, they are not tailored to the specific requirements of geotechnical construction, as we detail in \Cref{subsec:prev}. \citet{phoon2022unpacking} show that in order for the geotechnical engineering field to benefit from emerging digital technologies, these must be tailored to the unique characteristics of the field, where data is sparse, incomplete, of low quality and originates from multiple sources.

The objective of this work is to develop a Probabilistic Digital Twin (PDT) framework that can be used for geotechnical design and construction. Specifically, we show that to enable a framework that is scalable and can integrate the entire process, data should be divided into two types that affect the probabilistic digital twin at different time steps. The effectiveness of the PDT framework is demonstrated through an application to a highway foundation construction project, demonstrating its potential to improve decision-making and project outcomes in the face of uncertainties. After the PDT is created, the optimization of this decision-making process is compared with state-of-the-art optimization based on Monte Carlo Simulation.

The remainder of the paper is organized as follows. \Cref{sec:pdt} first reviews existing PDT approaches and then proposes a modified PDT framework to address the needs of AECOM systems and projects. The mathematical formulation of this framework follows in \Cref{sec:mat_description}. In \Cref{sec:geotech}, the specifics of each PDT component are discussed in the context of geotechnical design and construction. \Cref{sec:case_study} provides an analysis of the benefits, potential, and challenges of implementing a PDT approach, illustrated through a numerical investigation and the results in \Cref{sec:results}. This is followed by a discussion in \Cref{sec:discussion} and conclusion in \Cref{sec:conclusion}.

\section{Probabilistic digital twin concept}
\label{sec:pdt}

\subsection{Traditional Digital Twins}

As more industries began adopting the DT concept, there followed a series of definitions and interpretations \citep{brilakisborrman2019}. In the literature, various review articles are available that examine definitions, current approaches, challenges, and opportunities across different industries \citep[e.g.,][]{liu2021review, opoku2021digital, semeraro2021digital, singh2022applications}. Most definitions share three fundamental components of a digital twin: the physical entity, the digital replica, and the bidirectional communication between the two. \citet{Kritzinger2018} state that, compared to other digital representations, digital twins place a specific emphasis on the dual identification-control aspect. Accordingly, we adopt the definition of \citet{Kritzinger2018}, where virtual representation types of physical entities are categorized into three categories: (a) the \textit{Digital Model}, which involves manual modeling of the physical entity for visualization purposes with limited practical application; (b) the \textit{Digital Shadow}, where the virtual entity is continually updated through automated data streams from the physical entity, enabling its use in progress monitoring and quality control; (c) the \textit{Digital Twin}, which includes automated information exchange in both directions and is capable of accurately mirroring the state of the physical entity. The insights gained from modeling and predicting the behavior of the physical entity can be leveraged for operational decision-making.

\subsection{Prior Work in Probabilistic Digital Twins}
\label{subsec:prev}

\citet{kapteyn2021probabilistic} were among the first to formalize a probabilistic approach to digital twins, particularly for unmanned aerial vehicles. In their work, the digital twin is modeled as a probabilistic graphical model, using Bayesian methods for model updating and decision-making. This framework is scalable and can be extended to applications across multiple industries. For example, \citet{chaudhuri2023predictive} adapted this framework to the needs of risk-aware clinical decision-making, facilitating anticipatory, personalized tumor treatment that accounts for uncertainty. \citet{torzoni2024digital} extended the approach to structural health monitoring and management of civil engineering structures. They highlight the potential for assimilating structural response data using deep learning methods, enabling a change towards data-driven, predictive maintenance practices of bridges. However, these approaches primarily focus on behavioral aspects of systems, where physical states are observed indirectly through structural response measurements. We refer to this type of data as \textit{behavior data}. This data type is distinguished from \textit{property data}, which consist of direct measurements of the properties of the physical twin. This classification is detailed in \Cref{subsec:pdt_data}. In many industries, property data carry low uncertainties but in the construction industry - and particularly in geotechnical engineering - they are associated with large uncertainty. For example, borehole soundings are used to determine soil types and properties at the measurement locations. However, due to typically sparse measurements and the inherent variability of soils, significant uncertainties arise when extrapolating to unmeasured locations. A comprehensive PDT designed to support the full process of geotechnical design and construction must therefore differentiate between behavior and property data.

\citet{kochunas2021digital} demonstrated the potential of quantifying uncertainties in the DT approach for nuclear power systems, with a focus on understanding the sources of uncertainty in the modeling process and their propagation throughout the entire lifecycle. \citet{nath2022probabilistic} introduced the PDT concept for additive manufacturing, specifically for the laser powder bed fusion process. By incorporating measurement and model uncertainties, this approach enables early-stage design optimization and predictive maintenance. \citet{alibrandi2022risk} introduced the concept of a risk-informed digital twin to support sustainable and resilient engineering for urban communities. \citet{agrell2023optimal} provided a formal mathematical definition of a PDT and highlighted its potential for sequential decision-making under uncertainties. Despite these advancements, these approaches to PDTs are often tailored to specific applications, making their application to geotechnical design and construction unclear. They typically focus on a single phase of the PDT — design, construction, or operation — rather than addressing the entire lifecycle. As a result, their general applicability is limited, with each specific application requiring a customized PDT. This fragmented approach diminishes the potential benefits of integrating data and predictive behavior models into an unified framework.

To achieve a successful implementation of a PDT in geotechnical design and construction, the framework should ideally include several key components: (1) the creation of (3D) subsoil models from sparse property data (e.g. borehole soundings), (2) behavioral prediction of quantities of interest (e.g., settlement under load), (3) integration of established modeling methods to ensure trust, (4) initialization of historical data and expert knowledge at initial stages, (5) model updating for additional information, and (6) optimization of sequential decision-making under uncertainty.

In this work, we introduce a PDT framework that encompasses these key components. Like previous approaches, it extends the traditional digital twin to incorporate uncertainty quantification, enabling the use of Bayesian methods for data integration. However, it extends previous approaches, as it is specifically designed for scalability and differentiates between the two types of data. The PDT is demonstrated through a case study involving a highway foundation construction project, but it has broad applicability within AECOM industries. To the best of our knowledge, this research represents the PDT concept in the context of geotechnical design and construction.

\subsection{Probabilistic Digital Twin Framework}
\label{subsec:definition}

\begin{figure}[h]
\centering
\includegraphics[width=0.75\textwidth]{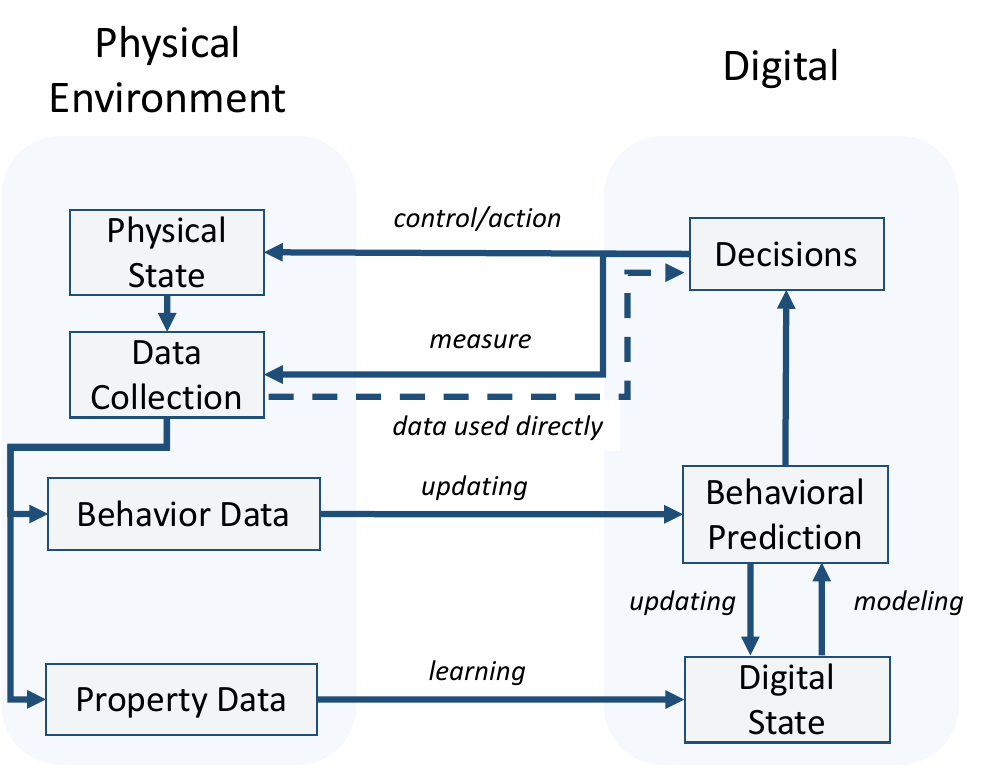}
\caption{Schematic flowchart for the proposed PDT framework}
\label{fig:dt_framework}
\end{figure}

In \Cref{fig:dt_framework}, the PDT framework is summarized. This framework encapsulates the \textbf{physical state}, representing the physical system and its interactions with the environment, from which \textbf{data} on properties and behaviors are obtained. In the \textit{learning} process, through statistical and mathematical modeling, \textbf{property data} is transformed into the \textbf{digital state}. This state captures all necessary attributes to model the physical state and is utilized in \textit{modeling} the \textbf{behavioral prediction}, which mirrors the behavior of the physical twin to predict future states. \textbf{Behavior data} is used to \textit{update} and calibrate the parameters of the behavioral prediction models, as well as the digital state. These predictions support \textit{decision optimization}, aiming to identify optimal \textbf{decisions} regarding \textit{controlling actions}, which alter the physical state, and decisions to \textit{collect additional information} to diminish model uncertainties.

In subsequent subsections, each component and process is discussed within the scope of construction engineering. We illustrate how each component relates to components of the traditional DT approach and how they are extended to accommodate uncertainties.

\subsubsection{Physical state}
\label{subsec:pdt_physical}

The physical state includes both the physical twin, which is the object of the PDT modeling process, and its interactions with the surrounding environment.

\subsubsection{Digital state}
\label{subsec:pdt_db}

At the beginning of the modeling process, a detailed requirements analysis should identify the quantities of interest, which are the critical parameters of the physical twin that should drive the decision optimization process. To predict the evolution of quantities of interest over time, appropriate models and their input parameters are required. Such an analysis is a complex task that necessitates extensive domain expertise, comprehensive knowledge of applicable methodologies, and the necessary level of detail for each project phase. As a rule, simpler models generally suffice for initial phases, whereas more detailed models are necessary for later stages. This model progression approach is comparable to the \textit{Level of Detail} used in disciplines like computer graphics and recognized in the AECOM industries under various terminologies \citep{abualdenien2022levels}. The PDT could leverage this concept to quantify model uncertainties correlated to the level of model detail.

Essential questions that should be addressed in the requirement analysis include: What is the object of the modeling effort? What are the quantities of interest, quantities of the physical twin? Which models are appropriate for this purpose? What properties are essential for these models to function effectively? What methods are available to acquire the necessary observations? What actions are possible to change the state of the physical twin? By what operational constraints are they affected? Among others, these questions form a robust foundation for a modeling strategy and result in a set of parameters that contain all important aspects of the physical twin, thereby enhancing the accuracy and utility of the PDT. A distinctive feature of the PDT approach is its ability to represent these parameters both deterministically and stochastically, allowing it to incorporate the inherent uncertainties and unknowns within complex systems.

\subsubsection{Data}
\label{subsec:pdt_data}

We distinguish two classes of data:

(1) \textbf{Property data} are direct observations of the attributes or properties of the physical twin (e.g., shear strength of soils or concrete). They are used to learn the parameters included in the digital state. The observations are obtained through intrusive methods (e.g., field tests, compressive strength tests, fatigue testing) or non-intrusive methods (e.g., geophysical, visual inspection of concrete, tomographic modeling). Sparse data availability is common, due to the damaging nature of excessive intrusive testing and the high costs for performing experiments.

(2) \textbf{Behavior data} is obtained by monitoring the behavior of the PT over time (e.g., soil settlement, building temperature, crack propagation in concrete). They do not provide direct insights into the state and attributes of the physical twin. Instead, they contain the monitored behavior over time and are used to calibrate predictive models and reduce their uncertainty.

Both classes of data are subject to observation uncertainty, which can be caused by limited measurement precision, faulty calibration of measurement devices, misreporting, among other reasons. This uncertainty is quantified by likelihood functions \citep{agrell2023optimal}.

\subsubsection{Learning the digital state}
\label{subsec:pdt_modeling}
To learn the digital state from property data commonly two steps are performed: 

(1) Where properties of interest cannot be measured directly, \textit{transformation} models are required to obtain them from property data. For example, in geotechnical engineering, empirical transformation models are used to categorize soil types (e.g., clay, sand, silt) based on mechanical properties derived from cone penetration test soundings \citep{Robertson2009}. In this step, uncertainties introduced due to the assumptions of the transformation model have to be accounted for \citep[e.g.,][]{Wang2016,Wang2018,wang2019bayesian}.

(2) \textit{Inference} of properties at unobserved locations is required due to the typical sparsity of available property data. In geotechnical engineering, Kriging interpolation, or Gaussian Process Regression (GPR), is commonly applied to simulate subsoil models \citep[e.g.,][]{Gong2020,yoshida2021estimation}. This probabilistic approach can quantify uncertainties introduced during modeling, which depend on the underlying assumptions and the quality and quantity of available data \citep{Rasmussen2005}.

The traditional DT approach is not capable of incorporating these uncertainties. Therefore, even when stochastic methods like GPR are used, the resulting uncertainties are frequently discarded, and only the expected values are utilized for predictions. In comparison, the PDT approach actively incorporates and propagates uncertainties through all modeling stages to enhance model robustness. By integrating stochastic learning methods, sparse data can be supplemented with information from similar regional projects or expert domain knowledge.

\subsubsection{Behavioral prediction modeling and updating}
\label{subsec:pdt_behavioral}

The predictive models within the DT framework aim to accurately simulate the behavior of their physical counterparts \citep{chaudhuri2023predictive}. Such models establish relationships between measurable physical attributes and the behavior of the physical twin \citep{agrell2023optimal}. By capturing these relationships, the models predict quantities of interest that inform subsequent decision optimization processes.

Traditionally, behavioral prediction models have been categorized into two main categories: physics-based models (e.g., finite element methods), which rely on an explicit representation of the physics of the system, and data-driven models (e.g., regression models, machine learning), where observation are directly interpreted to learn patterns and correlations. To overcome the limitations of each type, hybrid models were developed (e.g., physics-informed machine learning \citet{karniadakis2021physics}). They combine the interpretability and accuracy of physics-based models with the computational efficiency of data-driven models. Additionally, surrogate models are approximations of complex models designed to capture their essential features. Examples include artificial neural networks \citep{zhang2021application}, Gaussian Process models \citep{gramacy2020surrogates} or polynomial chaos expansion \citep{sudret2014polynomial}. The PDT uses Bayesian inference to update prediction models as new information becomes available. Bayesian inference is a statistical method that combines prior probability distributions with measurement likelihoods to estimate a posterior distribution. This approach can be used within the PDT framework since uncertain parameters are represented as random variables.

Typically, throughout the life cycle of the physical twin, various models are utilized to predict quantities of interest, each chosen based on the phase-specific requirements and objectives. Addressing this, the PDT framework is designed to be scalable, as it does not restrict the number of prediction models that can be integrated. The benefit of integrating multiple stochastic prediction models is their interconnection, which facilitates the propagation of uncertainties.

\subsubsection{Decision making}
\label{subsec:pdt_decisions}

\paragraph{Levels of automation}

Decision-making in AECOM industries is particularly challenging due to the societal impact and size of projects. In addition, personally liable, and decisions at every step have to be well-founded and explainable. The goal of the PDT framework is to actively manage uncertainties and provide support for decision-making in near-real-time. However, before achieving effective support for decision-making, trust must be established, and the accuracy of used models should be assured. Explainability of decisions is a crucial factor. For this reason, we envision a two-level adoption of the PDT approach:

(1) \textit{Semi-automated level}: In this stage, a hybrid human-computer decision-making approach is employed. The objective is to build trust and improve decision-making to a level replicating at least the accuracy of human counterparts. At this stage, the PDT serves as a supportive platform, offering decision recommendations to engineers who then choose how to proceed.

(2) \textit{Intelligent support for decision-making}: Achieving intelligent support for decision-making in the AECOM industry necessitates not only improved accuracy and increased trust but also significant technological advancements. When technological breakthroughs increase the degree of automation on site, the PDT framework is designed to automate and optimize support for the decision-making process. For example, in the future, automated, driver-free excavators could rely on the PDT to decide where and how much to dig. Simultaneously, the data collected by the sensors of the excavator could be used to update the digital state so that the system learns from its experience.

\paragraph{Types of decisions}
\label{subsub:decision_types}

Decisions in engineering are usually of two types: (1) \textit{information collection} $e$ to diminish epistemic uncertainties and refine model predictions and (2) \textit{controlling actions} $a$ aimed at enhancing system design or performance \citep{benjamin2014probability}. 

To facilitate good decision-making, actions are evaluated using behavioral prediction models to forecast their impact on the key quantities of interest. The output of the evaluation is a reward that is a decision metric used to objectively compare decisions. In the context of AECOM, potential rewards include reliability, cost, or environmental impact. Incorporating uncertainties in the PDT increases the computational demands, especially when decisions must account for both present and future state uncertainties. Optimization methods include Partially Observable Markov Decision Process (POMDP) solvers, Reinforcement Learning and Heuristic Strategy Optimization \citep{porta2005robot,roy2005finding,silver2010monte,papakonstantinou2018pomdp,andriotis2019managing,bismut2021optimal}.

\section{Mathematical model of the Probabilistic Digital Twin}
\label{sec:mat_description}

\begin{figure*}[h]
\centering
\includegraphics[width=0.9\textwidth]{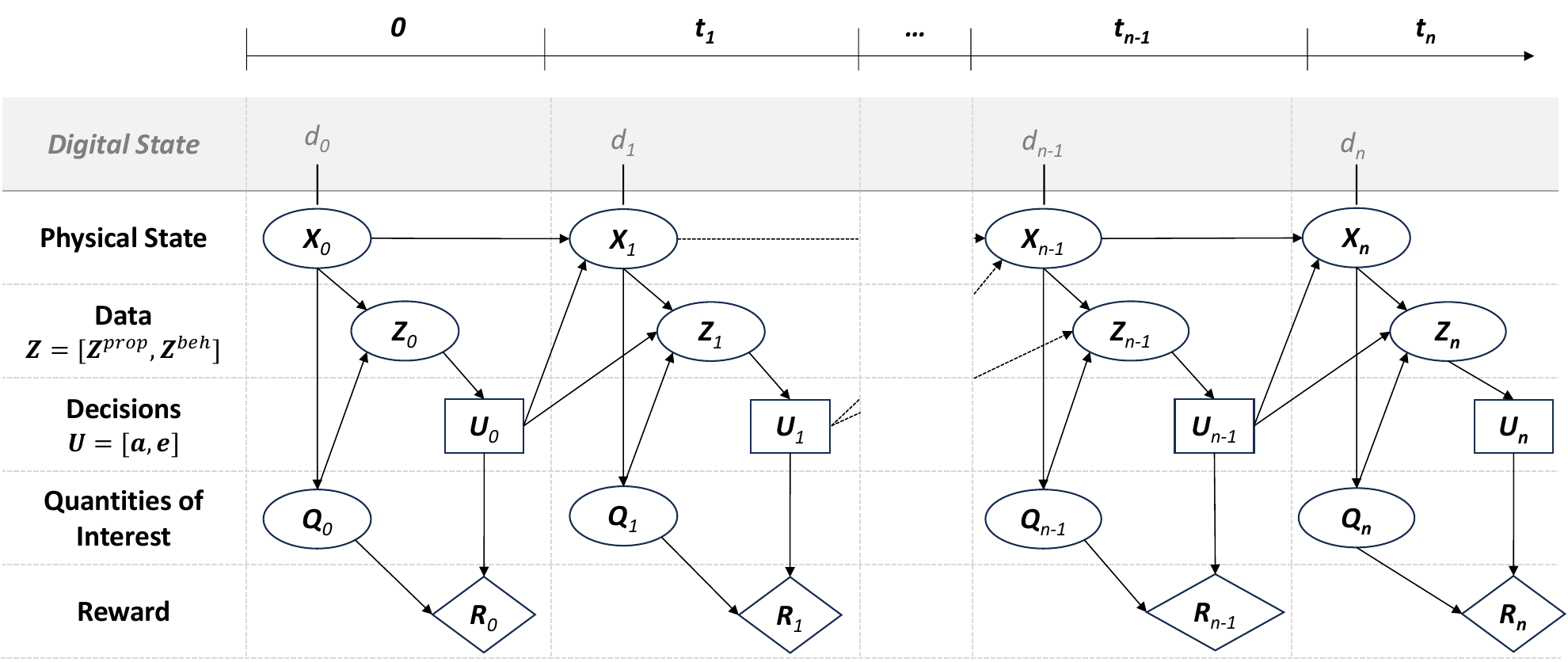}
\caption{The proposed PDT model is represented by an influence diagram to highlight the conditional dependencies of the individual components}
\label{fig:dt_if_diagram}
\end{figure*}

An influence diagram is utilized to model the PDT, shown in \Cref{fig:dt_if_diagram}. Introduced by \citet{shachter1986evaluating}, influence diagrams serve as a graphical tool to represent decision-making scenarios under uncertainty \citep{jensen2007bayesian,koller2009probabilistic}. Influence diagrams are acyclic-directed graphs in which round nodes represent random variables (RV), squared nodes represent decisions, and rhombus nodes represent utility functions. Directed arrows connect the nodes and specify the dependence structure among the RVs.

The PDT model incorporates the Markov assumption, which implies that the current state -- if known -- summarizes all the information contained in past states and data \citep{norris1998markov}. 
This allows for a simplified mathematical representation of belief states and their conditional dependencies across time. For applications where the Markov assumption does not hold a-priori, the model is still applicable through state space augmentation \citep{kitagawa1998self}.

The PDT model describes the relation between the physical state $\textbf{X}_t$, the data $\textbf{Z}$, which includes the property data $\textbf{Z}^{prop}$ and the behavioral data $\textbf{Z}^{beh}$, the quantities of interest $\textbf{Q}_t$, the decisions $\textbf{U}$ and the rewards $\textbf{R}_t$. As discussed in \Cref{subsub:decision_types}, the decisions are of two types $\textbf{U}=[\textbf{e}\textbf{,a}]$. The quantities of interest are the result of the behavioral prediction.

The state $\textbf{X}_t$ state is generally unknown and only represented probabilistically; it corresponds to the hidden state in a hidden Markov model \citep{koller2009probabilistic}. Therefore, what is known about $\textbf{X}_t$ is its (posterior) distribution. In the sequential decision-making under uncertainty literature, this distribution over the state is known as the \textit{belief} \citep{kochenderfer2015decision}. This belief is what we refer to as the digital state $d_t$ in this framework.

The digital state $d_t=p \left( \textbf{X}_t|\textbf{Z}_{0:t}=\textbf{z}_{0:t},\textbf{U}_{0:t-1}=\textbf{u}_{0:t-1} \right)$ is the distribution of $\textbf{X}_t$ conditional on all past and current observations $\textbf{z}_{0:t}$ and all past decisions $\textbf{u}_{0:t-1}$. It represents the knowledge of the physical state $\textbf{X}_t$ within the PDT. The digital state evolves dynamically as new data is obtained and decisions are made. Its transition dynamics and update with new data $\mathbf{Z}$ are given by
\begin{equation}
\label{eq:ifd_state}
\begin{split}
d_t  \propto &\sum_{\textbf{X}_{t-1}} \sum_{\textbf{Q}_{t}} 
    \underbrace{p \left( \textbf{X}_{t-1} | \textbf{Z}_{0:t-1}=\textbf{z}_{0:t-1}, \textbf{U}_{0:t-1}=\textbf{u}_{0:t-1} \right)}_{d_{t-1}} \times\\
    & \underbrace{p\left(\textbf{X}_t | \textbf{X}_{t-1}, \textbf{U}_{t-1}=\textbf{u}_{t-1} \right)}_{\phi^{transition}} \times \underbrace{p\left( \textbf{Z}_{t}=\textbf{z}_{t} | \textbf{X}_{t},\textbf{Q}_{t},\textbf{U}_{t-1}=\textbf{u}_{t-1} \right)}_{\phi^{data}} \times \underbrace{p\left( \textbf{Q}_{t}|\textbf{X}_{t} \right)}_{\phi^{pred}} \\
     &  = \sum_{\textbf{X}_{t-1}} \sum_{\textbf{Q}_{t}}  d_{t-1} \phi^{transition} \phi^{data} \phi^{pred},
\end{split}
\end{equation}
and has the following four components:
\begin{enumerate}
    \item $d_{t-1}= p \left( \textbf{X}_{t-1} | \textbf{Z}_{0:t-1}=\textbf{z}_{0:t-1}, \textbf{U}_{0:t-1}=\textbf{u}_{0:t-1} \right)$ is the digital state at the previous time step. It encapsulates the cumulative history up to time $t-1$. It is obtained by recursively applying \Cref{eq:ifd_state}, with an initial prior at $t=0$ that incorporates any pre-existing knowledge and expertise, as well as initial property data, following Section \ref{subsec:pdt_modeling}.
    
    \item $\phi^{transition}=p\left(\textbf{X}_t | \textbf{X}_{t-1}, \textbf{U}_{t-1}=\textbf{u}_{t-1} \right)$ is the state transition from $\textbf{X}_{t-1}$ to $\textbf{X}_t$ influenced by the controlling actions $\textbf{a}_{t-1}$.  

    \item $\phi^{data}= p \left( \textbf{Z}_{t}=\textbf{z}_{t} | \textbf{X}_{t},\textbf{Q}_{t},\textbf{U}_{t-1}=\textbf{u}_{t-1} \right)$ 
    is the likelihood of describing the data. It is conditional on $\textbf{e}_{t-1}$, the decision on what data should be collected. Given the distinction between the two types of data, this component can be divided into two parts: 
    \begin{equation}
     \phi^{data} = \phi^{prop}\phi^{beh},
    \end{equation}
    with $\phi^{prop}=p \left(\textbf{z}^{prop}_{t} | \textbf{X}_{t},\textbf{U}_{t-1}=\textbf{u}_{t-1} \right)$ being the likelihood of the property data and $\phi^{beh}=p \left(\textbf{z}^{beh}_{t} | \textbf{Q}_{t},\textbf{U}_{t-1}=\textbf{u}_{t-1} \right) $ the likelihood of the behavior data.

    \item $\phi^{pred}=p\left( \textbf{Q}_{t}|\textbf{X}_{t} \right)$ describes the dependency of the quantity of interest $\textbf{Q}_{t}$ on the state $\textbf{X}_t$.
\end{enumerate}

\begin{figure}[h]
\centering
\includegraphics[width=0.8\textwidth]{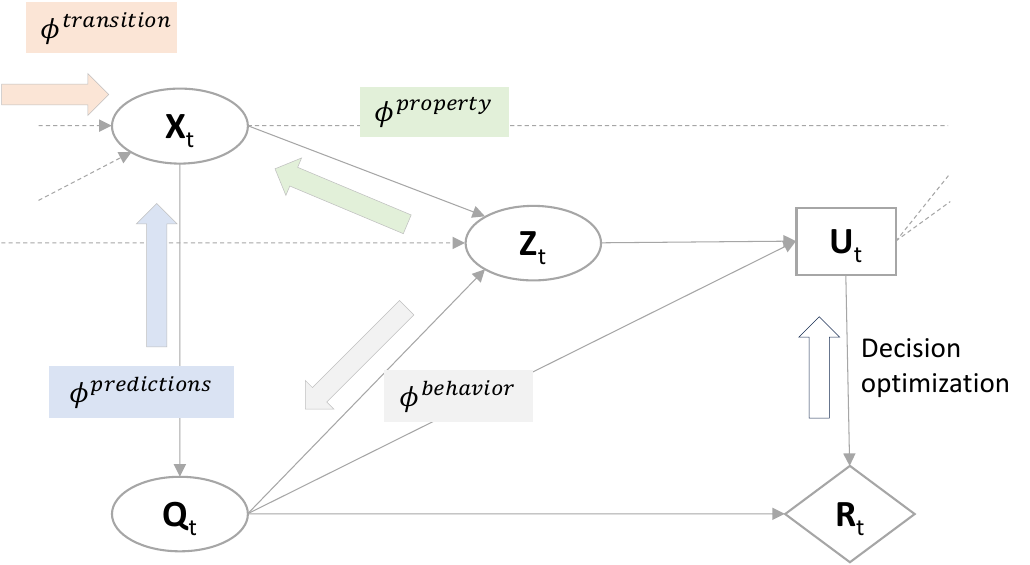}
\caption{Illustration of the transition and updating with new data $\textbf{z}_t= \left[ \textbf{z}_t^{prop}, \textbf{z}_t^{beh} \right]$ in one time step of the PDT}
\label{fig:dt_if_diagram_explained}
\end{figure}

The evolution and updating of the digital state according to \Cref{eq:ifd_state} results from the rules for probabilistic graphical models \citep{koller2009probabilistic}. \Cref{fig:dt_if_diagram_explained} is a graphical representation of this process, highlighting the opposite flow of information in the updating process compared to the causal dependencies of the underlying graphical model.

While \Cref{eq:ifd_state} implies that the physical state and the quantity of interest are discrete random variables, the PDT formulation is general. Continuous random variables can be included by replacing the summations in the equation with integrals. However, in practice,  solutions with straightforward numerical integration are not possible for high-dimensional problems. In such cases, sampling approaches can be used to create an approximation of $\mathbf{X}_t$ \citep[][]{russell2010artificial}.

One such approach is the particle filter (PF) approach \citealp{doucet2001introduction,doucet2009tutorial}, which is simple and has the ability to perform online updating \citep{kamariotis2023off}. It is a sequential importance sampling technique for approximating the posterior with weighted samples. 

The basic PF begins at $t=0$ by generating samples of $\mathbf{X}_0$ from the initial digital state (the prior distribution of $\mathbf{X}_0$). These samples are subsequently updated based on behavior data (settlement measurements) and actions (surcharge adjustments). At each time step, the Bayesian update is performed by evaluating $\phi^{beh}_k$ for every sample $k$ and computing normalized sample weights as:
\begin{equation}
    w_k = \frac{\phi^{beh}_k}{\sum_{k=1}^{n_s} \phi^{beh}_k}.
\end{equation}
Then, a resampling step is performed by randomly selecting (with replacement) $n_s$ samples according to the weights $w_k$.

One common issue with the basic PF approach is the sample degeneracy problem, in which most of the particles end up having weights close to zero after some updating steps. To overcome this issue, more advanced particle filter methods, or more generally sequential Monte Carlo methods, have been developed \citep{doucet2001sequential,cappe2007overview,chopin2020introduction, kamariotis2023off}. Alternatively, Kalman filter-based methods can be employed \citep{li2015kalman,song2020adaptive}.

\section{Probabilistic Digital Twin for geotechnical design and construction}
\label{sec:geotech}
\subsection{Introduction}
\label{subsec:geotech_intro}

Geotechnical design is concerned with the design of structures that interact with soils. Accurately characterizing soils is challenging, as soils are spatially varying and anisotropic due to their complex geological formation process. Although the physical causes of soil formation are deterministic and obey the laws of physics, it is currently impossible to fully understand how they combine. Furthermore, it is difficult to study their variation over time with incomplete knowledge, due to the impossibility of acquiring exhaustive subsoil property information. As a result, soil formation is usually assumed to be random \citep{Webster2000}. For the task of estimating geotechnical properties, \citet{Kulhawy2006} refer to this uncertainty as the \textit{inherent soil variability}.

In addition, three other sources of uncertainty exist in geotechnical engineering: measurement errors, modeling uncertainty, and statistical uncertainty \citep{Phoon1999}. Measurement errors emerge during data collection and are influenced by the equipment, methodologies, and personnel involved. Modeling uncertainties arise when measurement data is transformed into soil properties or quantities of interest, while statistical uncertainties are attributed to the scarcity of in-situ measurements and the methods used to extrapolate to unobserved locations.

To address the uncertainties, specific design approaches were adapted in geotechnical engineering. This includes the observational method, which was first introduced by \citet{peck1969advantages} and since has been incorporated into Eurocode 7. It ask for a continuous design review during construction with predefined contingency actions for deviations from acceptable behavior limits \citep{spross2017observational}. This aligns with the PDT approach, which offers a systematic framework for managing the complexities and uncertainties inherent in geotechnical engineering.

In addition to uncertainties, geotechnical construction of soil data requires the integration of diverse data sources, including boreholes, cone penetration test soundings, trenches, on-site tests, and laboratory-tested site-specific samples \citep{Zhang2018}. Technological advancements have increased the volume of geotechnical data collected, necessitating advanced information management tools for effective data integration and decision-making support \citep{Chandler2011,Zhou2013,Phoon2019}.

\subsection{Application to an embankment}
\label{subsec:geotech_case_study}

\begin{figure}[h!]
\centering
\includegraphics[scale=1]{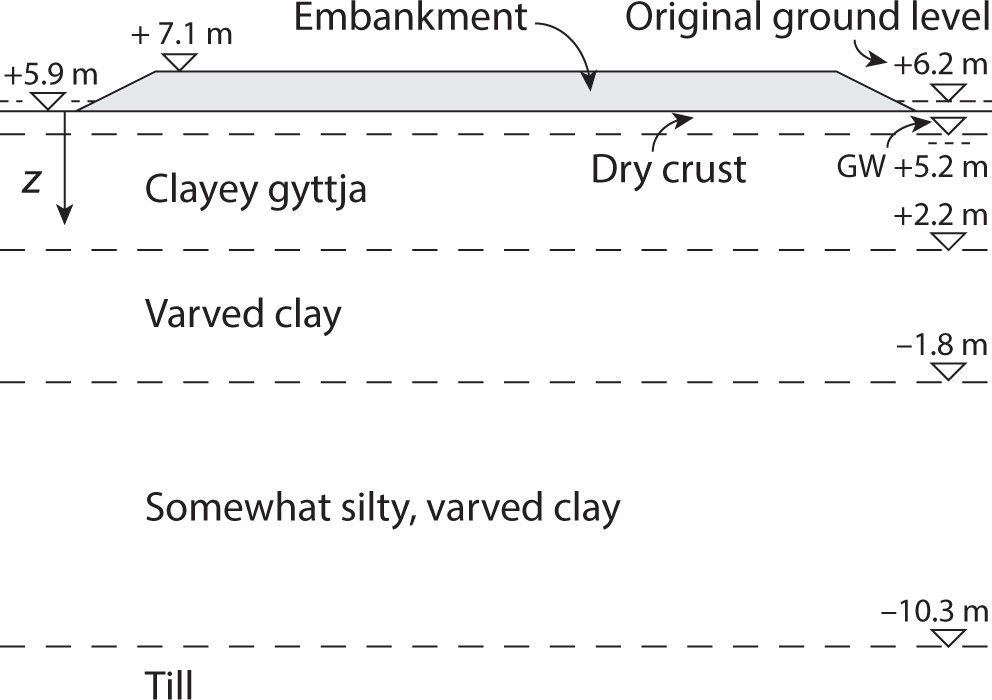}
\caption{Cross-section of the soil under the planned embankment \citep[from ][\href{https://creativecommons.org/licenses/by/4.0/}{CC-BY-4.0}]{spross2021probabilistic}}
\label{fig:soil_embankment}
\end{figure}

To demonstrate the PDT framework in geotechnical engineering, the construction of a highway on clayey soil is considered (see \Cref{fig:soil_embankment}) following \citet{spross2021probabilistic} and \citet{bismut2023optimal}. A consolidation process, which causes settlements, is started by loading the clayey soil with an embankment. The consolidation converges towards an equilibrium and is dependent on the load size. Before the road on top of the embankment is constructed, the consolidation should have converged to avoid any damage to the road. To speed up the consolidation process, prefabricated vertical drains (PVDs) are installed to facilitate water drainage, and the embankment is preloaded with a surcharge. The task of engineers is to find the most cost-efficient surcharge and PVD design, which ensures that the consolidation process is finalized within a predetermined timeframe $t_{max}$. This is a challenging task due to the large uncertainties for predicting the time and magnitude of the long-term settlement.

\citet{spross2021probabilistic} developed a probabilistic model that describes the settlement of an embankment on top of PVDs and loaded with a surcharge over time. It is a physics-based behavioral model that is calibrated to site-specific conditions using measured data. Building on this, \citet{bismut2023optimal} apply a risk-based framework for sequential decision-making under uncertainties to this task.

In this paper, we elaborate on how the PDT approach can be used to improve predictions and decision-making for the embankment problem. To enable this, the behavior prediction model of the surcharge is extended with Bayesian inference capabilities to integrate settlement measurements. In the following, the components of the PDT are introduced.

\subsubsection{Physical state}
\label{subsec:geotech_ps}

The example application is based on the construction of the \textit{Highway 73} in southern Stockholm. It focuses on a $550\,m$ road embankment constructed over a $0.3\,m$ dry crust atop $15.5\,m$ of soft clayey soil, with a targeted embankment height of $1.2\,m$. The geotechnical investigation data is also taken from this project. As it is not the focus here, the PVD design is assumed to be fixed. The design details are taken from \citet{spross2021probabilistic}.

The physical state is described by the vector 
\begin{equation*}
\label{eq:geotech_ds}
    \resizebox{.78\textwidth}{!}{$
	\mathbf{X}_t = \left[\mathbf{\alpha},  \Delta^{sur}(t),\underbrace{\sigma'_L, \sigma'_c, \gamma_{cl}, \gamma_{emb}, M_0, M_L, w_N, c_v, c_h}_{\textbf{X}_{soil}}, S_{\infty}\left(t\right), U(t)\right]$}
\end{equation*}
where $\mathbf{\alpha}$ are deterministic geometric boundary conditions of the embankment defined at the beginning of the project; $\Delta^{sur}(t)$ indicates the surcharge height which can change over time $t$; $\textbf{X}_{soil}$ are the relevant soil properties required for the behavior prediction models used in this problem; and the long-term settlement $S_{\infty}(t)$ and the degree of consolidation $U(t)$ are quantities predicted by the behavioral models and required for deriving the quantities of interest. An in-depth explanation of the model is provided in \Cref{subsec:geotech_behavior}.

In this application, $\phi^{transition}$ is influenced by the decision made in the previous step regarding adjustments to the surcharge height $\Delta^{sur}(t)$ and by parameters that change over time. Specifically, $\Delta^{sur}(t)$ affects the long-term settlement $S_{\infty}\left(t\right)$. The degree of consolidation $U(t)$ evolves over time and is mostly dependent on the PVDs.

\subsubsection{Data}
\label{subsec:geotech_data}
Following the PDT framework, we distinguish between the property data and the behavior data. 

\paragraph{Property data}\mbox{}

Creating a digital representation of subsoil conditions requires soil property data, typically acquired through intrusive and non-intrusive tests. For example, borehole soundings and cone penetration tests are commonly used to classify soil types and ascertain mechanical properties. Due to the costly and intrusive nature of such tests, data availability is often limited.

In this case study, samples of soil properties were collected at different depths from a single location on the embankment, identified to be problematic by engineers. These samples are used to learn the initial model, which is presented in \Cref{subsec:geotech_modelling}. During the construction process, no additional property data is collected to learn the PDT, resulting in $\phi^{prop}=1$ for $t>0$ for this investigation.

\begin{table*}[!h]
\caption{Geotechnical parameters modeled in the case study \citep{spross2021probabilistic} \label{tab:geotechnical_parameters}}
\small
\centering
\begin{adjustbox}{width=\textwidth}
\begin{tabular}{l|c|l}
\hline
Parameter & Symbol & Distribution \\
\hline
Limit pressure towards increasing modulus & $\sigma'_L$ & Lognormal (infered from 9 samples)\\
Preconsolidation pressure & $\sigma'_c$ & Lognormal (infered from 9 samples)\\
Unit weight of clay & $\gamma_{cl}$ & Lognormal (infered from 7 samples)\\
Unit weight of the embankment & $\gamma_{emb}$ & Lognormal$(\mu=20.8\,kN/m^3, \sigma^2=5\%\mu)$ \\
Modulus for $\sigma' \leq \sigma'_c$ & $M_0$ & Lognormal (infered from 9 samples)\\
Modulus for $\sigma'_c < \sigma' \leq \sigma'_L$ & $M_L$ & Lognormal (infered from 7 samples)\\
Natural water content & $w_N$ & Lognormal (infered from 7 samples)\\
Vertical consolidation coefficient & $c_v$ & Lognormal $(\mu=0.2\,m^2/year,\sigma^2=50\%\mu)$ \\
Horizontal consolidation coefficient & $c_h$ & Lognormal$(\mu=2.5*0.2\,m^2/year,\sigma^2=50\%\mu)$ \\
\end{tabular} 
\end{adjustbox}
\end{table*}

\paragraph{Behavior data}\mbox{}

Behavioral data in geotechnical engineering is crucial for validating and calibrating behavioral models and updating the digital state. In this case study, behavioral data is obtained from weekly measurements of the settlement $z_{s}(t)$ observed in the preloaded embankment. The associated measurement error is $\varepsilon$, with probability density function $f_\varepsilon$. The corresponding likelihood function is 
\begin{equation}
    \phi^{beh}=f_\varepsilon \left(z_{s}(t)-s(t) \right).
\end{equation}
$\varepsilon$ follows a normal distribution with mean zero and standard deviation $\sigma_\varepsilon$. In the subsequent numerical investigation, three distinct scenarios with varying $\sigma_\varepsilon$ are analyzed to assess the impact of this parameter on the results.

\subsubsection{Learning the initial digital state}
\label{subsec:geotech_modelling}

As illustrated in the PDT framework in \Cref{fig:dt_framework}, property data is taken as direct input in the modeling stage to create an initial digital model $d_0$. Following the steps outlined in \Cref{subsec:pdt_modeling}, \textit{transformation} and \textit{inference} are required at this stage.

For the embankment problem, \citet{spross2021probabilistic} describe the soil properties $\textbf{X}_{soil}$ as RVs. The core modeling assumption is that soil properties vary only with depth, adopting a 1D soil model perspective. Regression models are used to learn the distribution of soil properties over depth, which is the equivalent of learning the initial digital state $\textbf{X}_{soil}$ are summarized in \Cref{tab:geotechnical_parameters}.

\subsubsection{Behavioral prediction modeling}
\label{subsec:geotech_behavior}

The quantities of interest $\textbf{Q}(t)$ for this problem are the settlement $S(t)$ and overconsolidation ratio $OCR(t)$ over time.

\begin{figure*}[h!]
\centering
\includegraphics[width=0.5\textwidth]{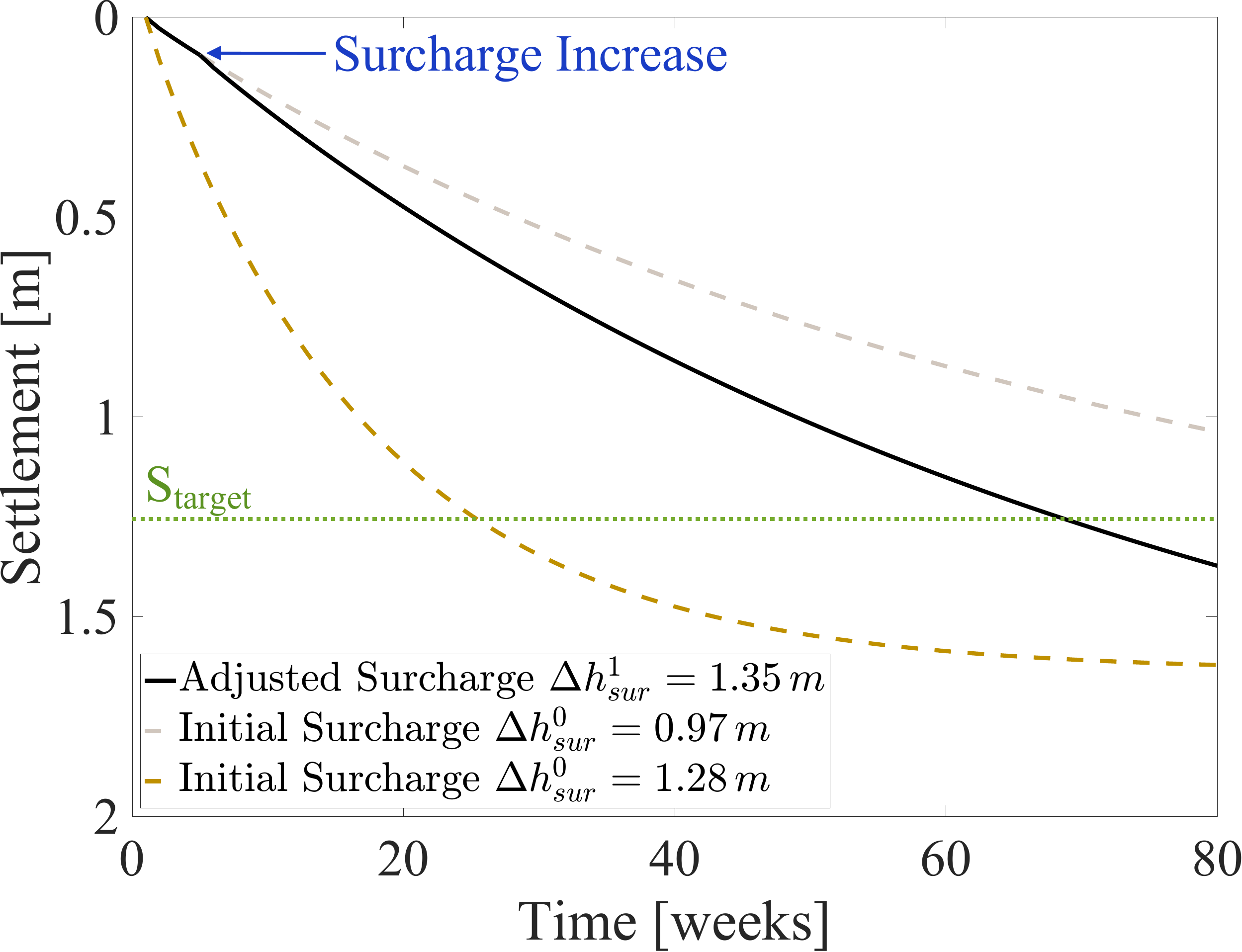}%
\includegraphics[width=0.5\textwidth]{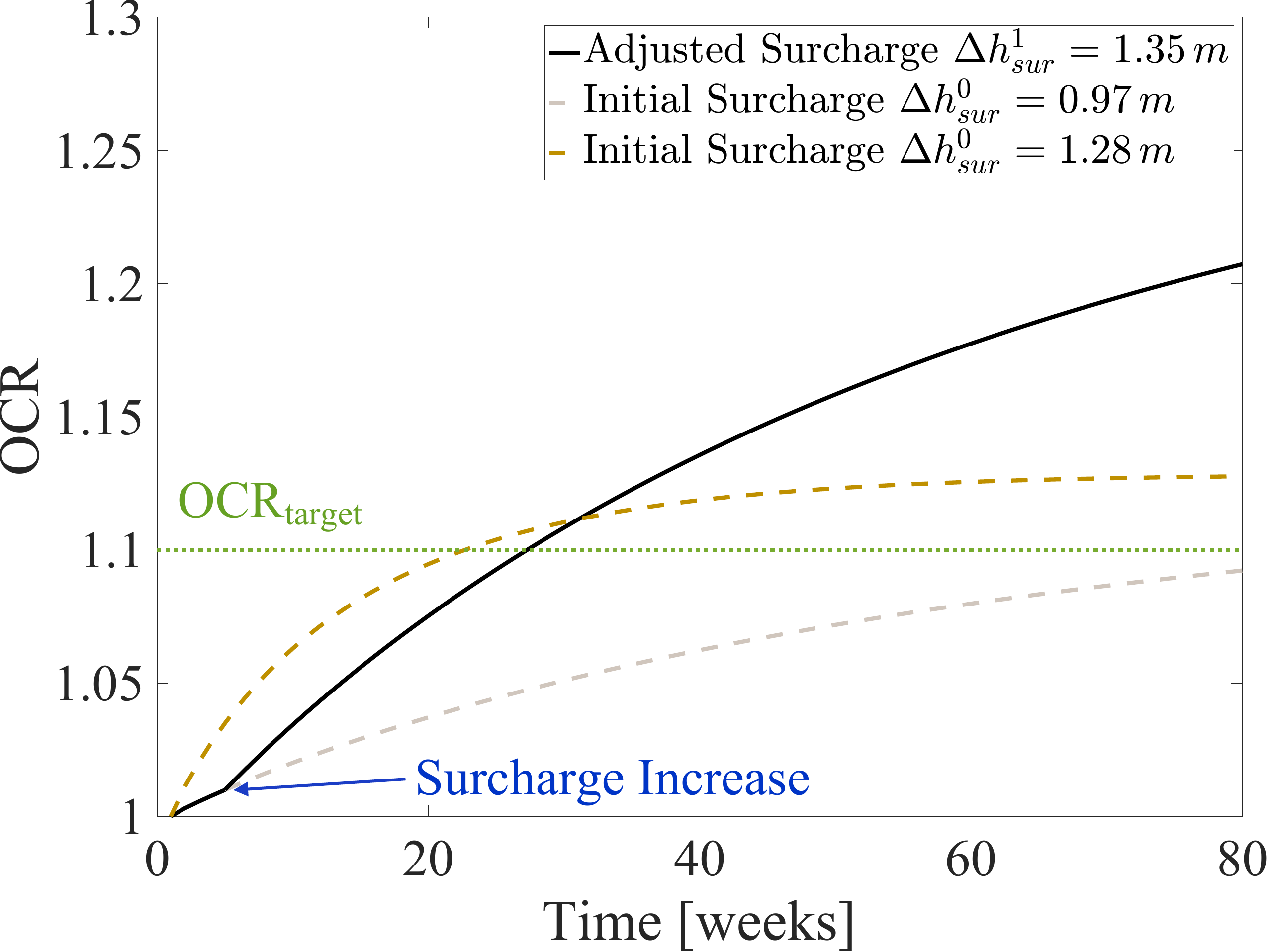}
\caption{Three example trajectories obtained from the geotechnical model for both a) settlement and b) overconsolidation ratio over time to demonstrate the impact of adjusting the surcharge on both parameters. The grey-dashed line is a trajectory where the settlement target $s_{target}$ is not met within the maximum project time $t_{max}=72 [weeks]$. Increasing the surcharge height, results in the black line trajectory where the requirement is successfully achieved. The dashed brown line shows a trajectory where the $s_{target}$ requirement is met without the need for any surcharge height adjustments}
\label{fig:stmax_ocrtmax}
\end{figure*}

Geotechnical engineering commonly employs deterministic, physics-based models to predict soil behavior. For the embankment scenario, \citet{spross2021probabilistic} utilize a traditional deterministic model for predicting the consolidation of embankments on top of PVDs and preloaded with a surcharge. The model takes as input the geotechnical stochastic parameters in $\mathbf{X}_t$ and provides a prediction of the consolidation over time. \citet{bismut2023optimal} extended the model to be able to account for the effect of adjusting the surcharge height during the preloading phase. This allows for more complex preloading strategies, as surcharge heights can be adapted when measurements indicate a low probability of reaching requirements. Example trajectories obtained with the model are depicted in \Cref{fig:stmax_ocrtmax}.

The probabilistic model is used to predict two parameters: (1) $S_\infty(t)$ the predicted primary long-term settlement for $t \rightarrow \infty$, which is dependent on the loading capacity of soil and the preloading strategy of the embankment with

\begin{equation}\label{Eq:Sinf}
S_{\infty}(t)=\sum_{i=1}^{l}b_{cl,i}\Delta\epsilon_{i}\left(\Delta\sigma(t)\right).
\end{equation}
It is the sum over the settlement of each layer $i$, which is given by the product between the clay layer thickness $b_{cl,i}$ and the increase in strain $\Delta\epsilon_i$ due to the load $\Delta\sigma(t)$ at time $t$. The load is dependent on the surcharge height $h_t$.

(2) $U(t)$, the spatially averaged degree of consolidation over time $t$, which is given as

\begin{equation}
    \label{eq:consolidation_at_t}
    U(t) = 1 - \left[ 1 - U_v(t)\right] \left[ 1- U_h(t) \right]
\end{equation}
where $U_v(t)$ and $U_h(t)$ are the vertical and horizontal consolidation component. They can be obtained from Terzaghi's consolidation theory and Hansbo's analytical PVD model \citep{spross2021probabilistic}.

The distribution of the settlement over time $S(t)$ is obtained as

\begin{equation}
    \label{eq:S_t}
S(t)=\begin{cases}
	U(t)S_{\infty}(t), & \text{for $0 \leq t<t_{add}$},\\
	U(t-t_{shift})S_{\infty}(t) & \text{for $t \geq t_{add}$}.
\end{cases}
\end{equation}
The model is capable of considering one surcharge increment $\Delta\sigma_{add}$ at time $t_{add}$. This results in the total embankment load 

\begin{align}
    \Delta\sigma(t)=\Delta\sigma_{emb}+\Delta\sigma_{sur}+\Delta\sigma_{add} &&  \text{for } t>t_{add}.
\end{align}
The parameter $t_{shift}$ is required to reflect the accelerated consolidation process due to an increased load, resulting in a larger $S_\infty$. For details on how to obtain it, we refer to \citet{bismut2023optimal}.

The overconsolidation ratio $OCR$ at time $t$ is given as

\begin{equation}
    \label{eq:OCR_t}
OCR(t)=\frac{\sigma'_{0}+U(t)\Delta\sigma_{sur}+\Delta U(t) \Delta\sigma_{add}}{\sigma'_{0}+U(t)\Delta\sigma_{emb}},
\end{equation}
and describes the ratio between preconsolidation stress due to the embankment load $\Delta\sigma_{emb}$ and current stress due to the added preloading $\Delta\sigma_{sur}+\Delta\sigma_{add}$. $\Delta U(t)$ is again required to account for the accelerated consolidation process due to the load increase. We refer to \citep{bismut2023optimal} for a detailed explanation of how it is calculated.

Following the above model, the quantities of interest in $\textbf{Q}(t)$ are a deterministic function of $\textbf{X}_t$, $\textbf{Q}(t)=q(\textbf{X}_t)$. Therefore, $\phi^{pred}$ is expressed through the Dirac delta function as $\phi^{pred}=\delta\left(\textbf{Q}_t-q(\textbf{X}_t)\right)$.

\subsubsection{Behavioral prediction updating}
\label{sec:bayesian_learning}

In this study, the basic particle filter (see \Cref{sec:mat_description}) is used to update the distribution of the settlement over time with weekly settlement measurements described in \Cref{subsec:geotech_data}. As expected, we also observe the sample degeneracy problem during the updating process. However, as we show in \Cref{sec:case_study}, the accuracy of the approach in this case suffices for improving the decision-making process.

\begin{figure}[!h]
\centering
\includegraphics[scale=0.24]{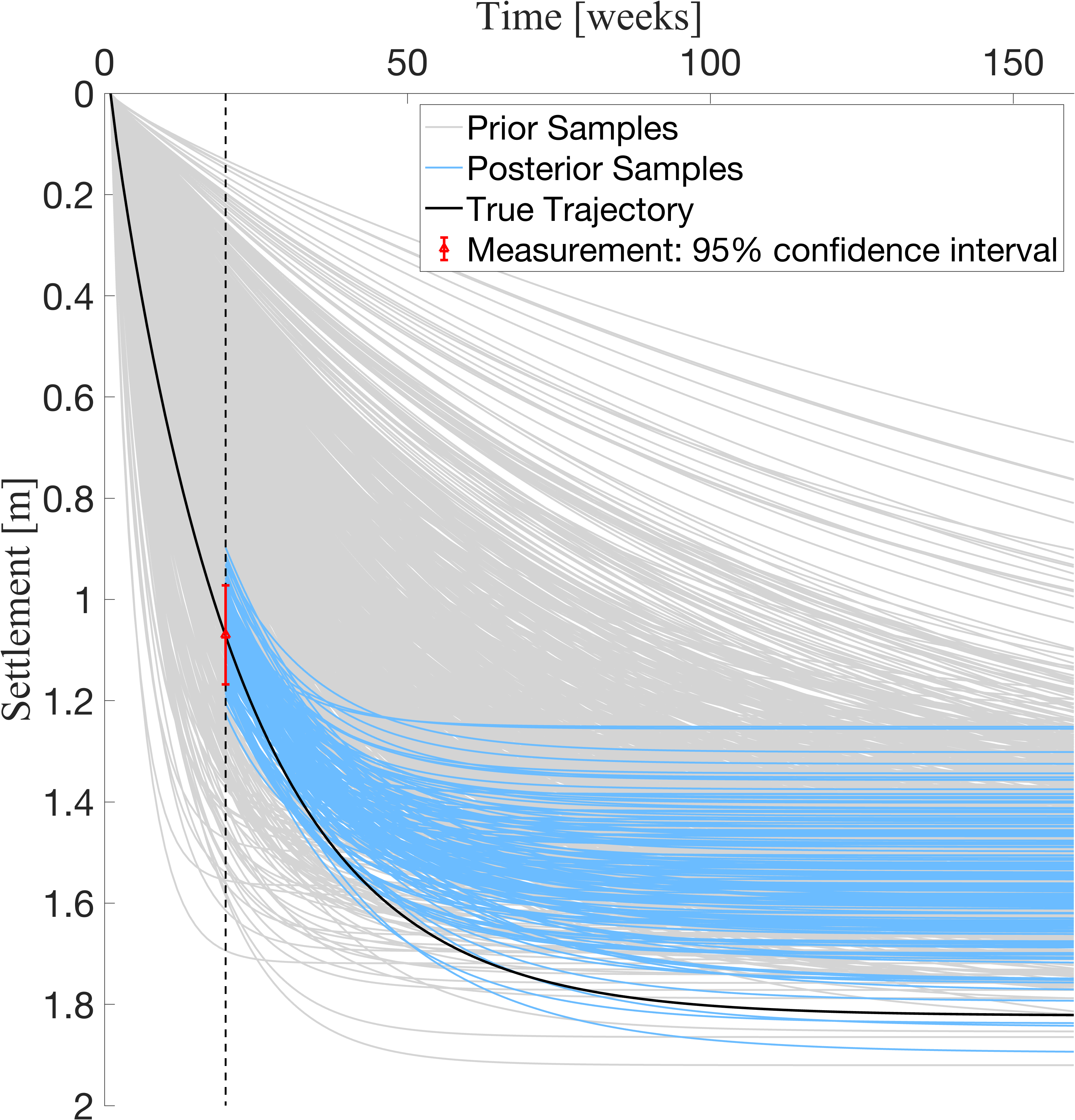}%
\includegraphics[scale=0.24]{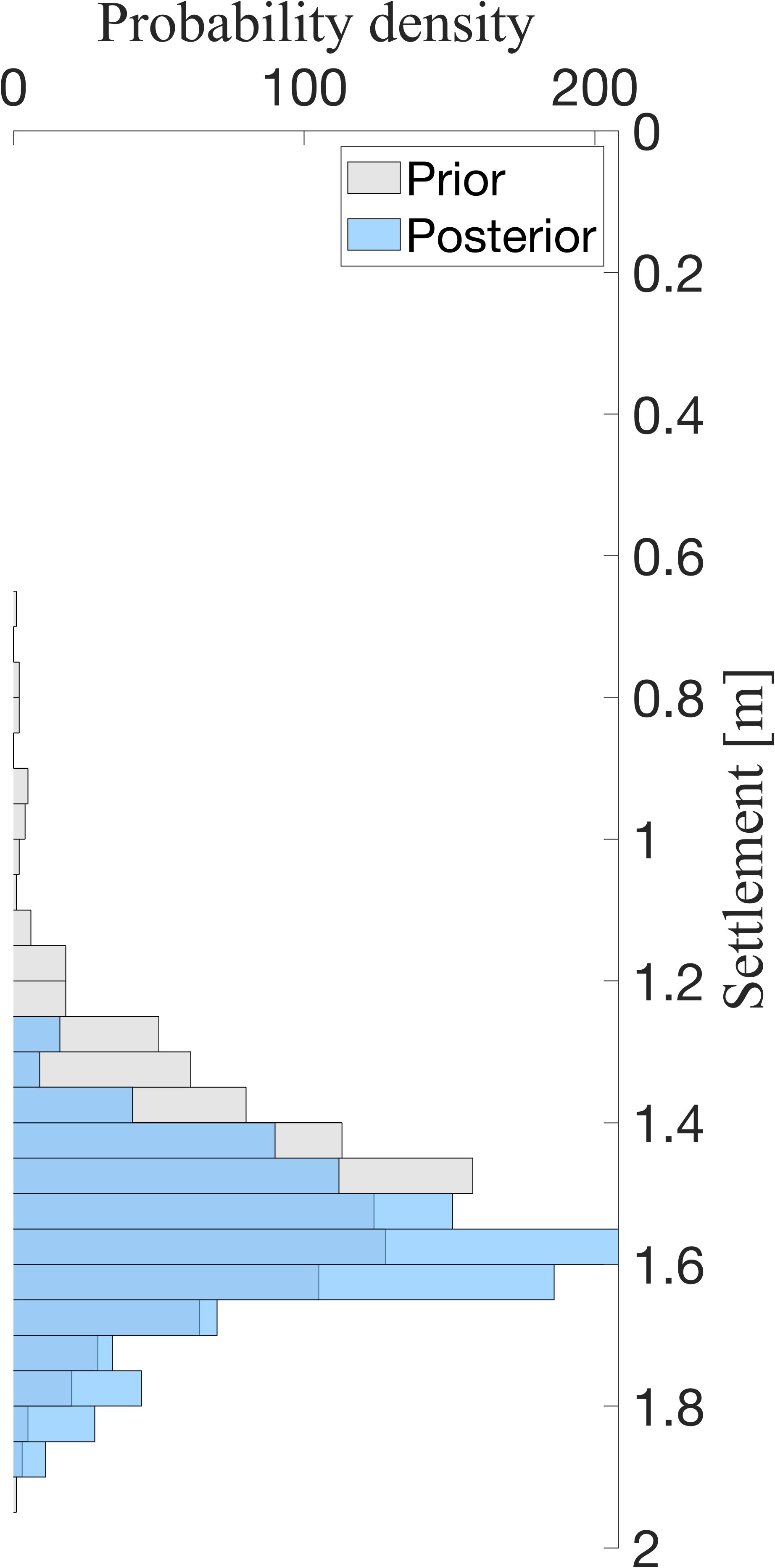}
\caption{$10 000$ trajectory samples obtained with the probabilistic model for an initial surcharge $h_0=0.9m$. On the left side, their evolution over time is illustrated. The posterior (highlighted in blue) is obtained for a measurement $z_s(t=20 weeks)=1.08m$. Measurement errors are modeled as a standard normal distribution $\sigma_\varepsilon = 5\,cm$ and added to the measurement. The histogram on the right illustrates both the prior and posterior of the distribution of $S$ at $t=160$}
\label{fig:posterior_evidence}
\end{figure}

In \Cref{fig:posterior_evidence} the prior and posterior of $S(t)$ obtained with particle filtering for a measurement at $t=20 [weeks]$ is illustrated.

\subsubsection{Decision making}
\label{subsec:geotech_decision making}

\paragraph{Decisions}

This application does not involve decisions $e_t$ to collect additional information, as a decision to collect weekly settlement measurements is taken a-priori.

Decisions are taken on the surcharge height. The first decision at time zero is the initial surcharge height $a_0=[h_0]$. 
In subsequent time steps, the decision alternatives are $a_t=\left[ \text{do nothing, adjust surcharge height by $h_1$}\right]$. 

\paragraph{Requirements}
For the embankment problem, \citet{spross2021probabilistic} specify requirements for the settlement and OCR at the time of unloading $t_{max}$. First,

\begin{equation}
    S(t_{max}) \leq s_{target},
\end{equation}
where $s_{target}$ represents the settlement threshold. This criterion ensures that residual primary consolidation remains within acceptable serviceability limits after unloading. This requirement is considered crucial and the embankment can only be unloaded when it is fulfilled. Not meeting this requirement at $t_{max}$ results in project delay and penalties.

The OCR requirement aims to prevent creep settlements after the road is taken into service:

\begin{equation}
    OCR(t_{max}) \geq OCR_{target}
\end{equation}
with $OCR_{target}=1.10$ as specified in the general technical requirements and guidance for geotechnical works issued by the Swedish Transport Administration \citep{spross2021probabilistic}. Failing to meet the OCR criterion can deteriorate road quality and reduce its lifespan, leading to bumps in the road, cracks and accelerated wear.

\paragraph{Cost function}
Following \citet{bismut2023optimal}, the cost function is
\begin{equation}
\label{eq:general_cost_function}
    C_{tot} = \sum_i C_{sur,i} + C_{delay} + C_{OCR},
\end{equation}
where $C_{sur,i}$ quantifies the costs of the surcharge (e.g., material, equipment and labor costs), with $i=1$ representing the initial construction costs and $i=2$ the costs for increasing the surcharge; $C_{delay}$ quantifies the penalty if $s_{target}$ is not reached in time; and $C_{OCR}$ quantifies the penalty of not reaching $OCR_{target}$ at time of unloading. For the full details of the cost function parameters, we refer to \citep{bismut2023optimal}.

In the PDT, the cost function is equivalent to the negative reward $\textbf{R}_t$ and the components of the summation in \Cref{eq:general_cost_function} are a function of $\mathbf{X}_{0:t{max}}$ and $\mathbf{a}_{0:t{max}}$. $C_{sur,i}$ depends on the surcharge decision $a_t$, whereas $C_{delay}$ and $C_{OCR}$ depend on quantities of interest at time $t_{max}$, i.e., $S(t_{max})$ and $OCR(t_{max})$. To make this dependence explicit, we use the notation $C_{tot}(\mathbf{X}_{0:t{max}},\mathbf{a}_{0:t{max}})$ in the following.

\paragraph{Optimal decision}

The objective of the decision optimization is to find the sequence of actions, i.e., surcharge loading that results in the lowest expected cost. The actions are defined by a decision strategy $\mathcal{S}$, which consists of a set of policies $\pi_t$ that define the action $a_t$ to take at time $t$ in function of the current digital state $d_t$, i.e., $a_t = \pi_t(d_t)$. The goal is to find the optimal strategy.

A heuristic approach is chosen to solve this decision problem \citep{bismut2021optimal}. It reduces the complexity of the problem by describing the strategy $\mathcal{S}$ through a parametric function $\mathcal{S}_{\mathbf{w}}$, where $\mathbf{w}=[w_1,w_2,...w_n]$ are the heuristic parameters.

The optimal heuristic strategy for a given heuristic function is defined by parameters
\begin{equation}
\label{eq:opt_heuristic}
   \mathbf{w}_{opt}=\arg \min_\mathbf{w} \mathbb{E} \left[C_{tot}(\mathbf{X}_{0:t{max}},S_{\mathbf{w}}(\mathbf{D}_{0:t{max}})) \right]
\end{equation}
and aims to find the minimal expected cost.

Monte Carlo simulation is used to obtain an approximation of the objective function of \Cref{eq:opt_heuristic}. $n_{MC}$ sample trajectories for a given surcharge height are simulated using the probabilistic geotechnical model. The MC approximation for a given set of heuristic parameters $\mathbf{w}$ is given as 
\begin{equation}
    \mathbb{E} \left[C_{tot}(\mathbf{X}_{0:t{max}},S_{\mathbf{w}}(\mathbf{D}_{0:t{max}})) \right] \approx \frac{1}{n_{MC}} \sum_{k=1}^{n_{MC}} 
    C_{tot}(\mathbf{X}^{(k)}_{0:t{max}},S_{\mathbf{w}}(\mathbf{D}^{(k)}_{0:t{max}}))   
\end{equation}
wherein $\mathbf{X}^{(k)}_{0:t{max}}$ are the $k=1:n_{MC}$ sample trajectories of the true state and $\mathbf{D}^{(k)}_{0:t{max}}$ are the corresponding samples of the digital state.

To solve \Cref{eq:opt_heuristic}, the cross-entropy (CE) optimization is used \citep{Kroese_et_al_06, Bismut_et_al_22}. It is an optimization method tailored to noisy optimization problems. For details on the employed CE algorithm, we refer to \citet{bismut2023optimal}.

\subsubsection{Summary}

\begin{figure*}[h]
\centering
\includegraphics[width=\textwidth]{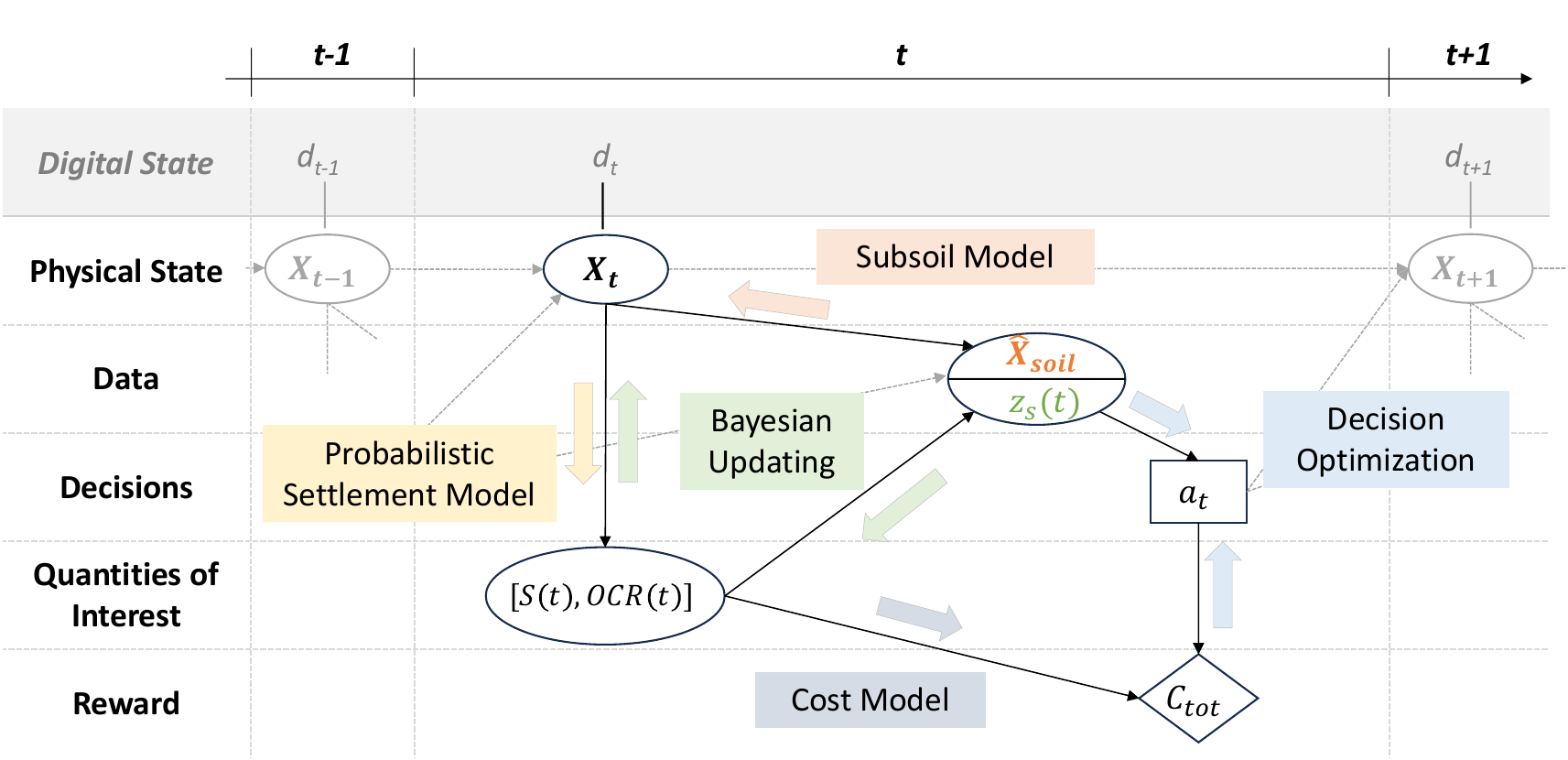}
\caption{The PDT influence diagram of \Cref{fig:dt_if_diagram}, adapted to the considered application case}
\label{fig:if_diagram_geotech}
\end{figure*}

\Cref{fig:if_diagram_geotech} shows the influence diagram of the general PDT tailored to this specific application. 
Here, the physical state $\textbf{X}_t$ includes all the parameters identified as necessary to describe the embankment and its settlement over time. The state is learned in the initial stage from collected property data samples of $\textbf{X}_{soil}$. These soil properties are the input of the probabilistic settlement model by \citet{spross2021probabilistic} for the prediction of the quantities of interest $\textbf{Q}(t)=\left[S(t),OCR(t)\right]$. Behavioral data for this problem consists of measurements of the settlement $\textbf{z}_s(t)$, which are used to update the physical state $\textbf{X}_t$ to enhance the prediction accuracy of the quantities of interest. Decisions on the surcharge height $a_t$ are based on the posterior predictions and directly affect the total expected cost $C_{tot}$. Heuristics for decision optimization of the preloading strategy are employed to identify actions that minimize the costs over the lifecycle of the embankment.

\section{Numerical investigation}
\label{sec:case_study}

\subsection{Model setup}

The numerical investigation is performed for the embankment of a highway section constructed in southern Stockholm, Sweden, introduced in \Cref{subsec:geotech_ps}. The potential of the PDT, illustrated in \Cref{fig:if_diagram_geotech}, for the optimization of preloading strategies is investigated. The settlement target $s_{target}=1.27\,m$ is according to \citet{spross2021probabilistic}. The cost function parameters are adopted from \citet{bismut2023optimal}.

\subsection{Decision heuristic}

In this study, we introduce a heuristic $\mathcal{S}_{\mathbf{w}}$ that leverages the PDT to improve decision-making for the embankment construction. This heuristic is later compared to heuristics established in previous work\citep{bismut2023optimal, cotoarba2023data}, which are based on data rather than model predictions.

\begin{figure*}[h!]
\centering
\includegraphics[width=\textwidth]{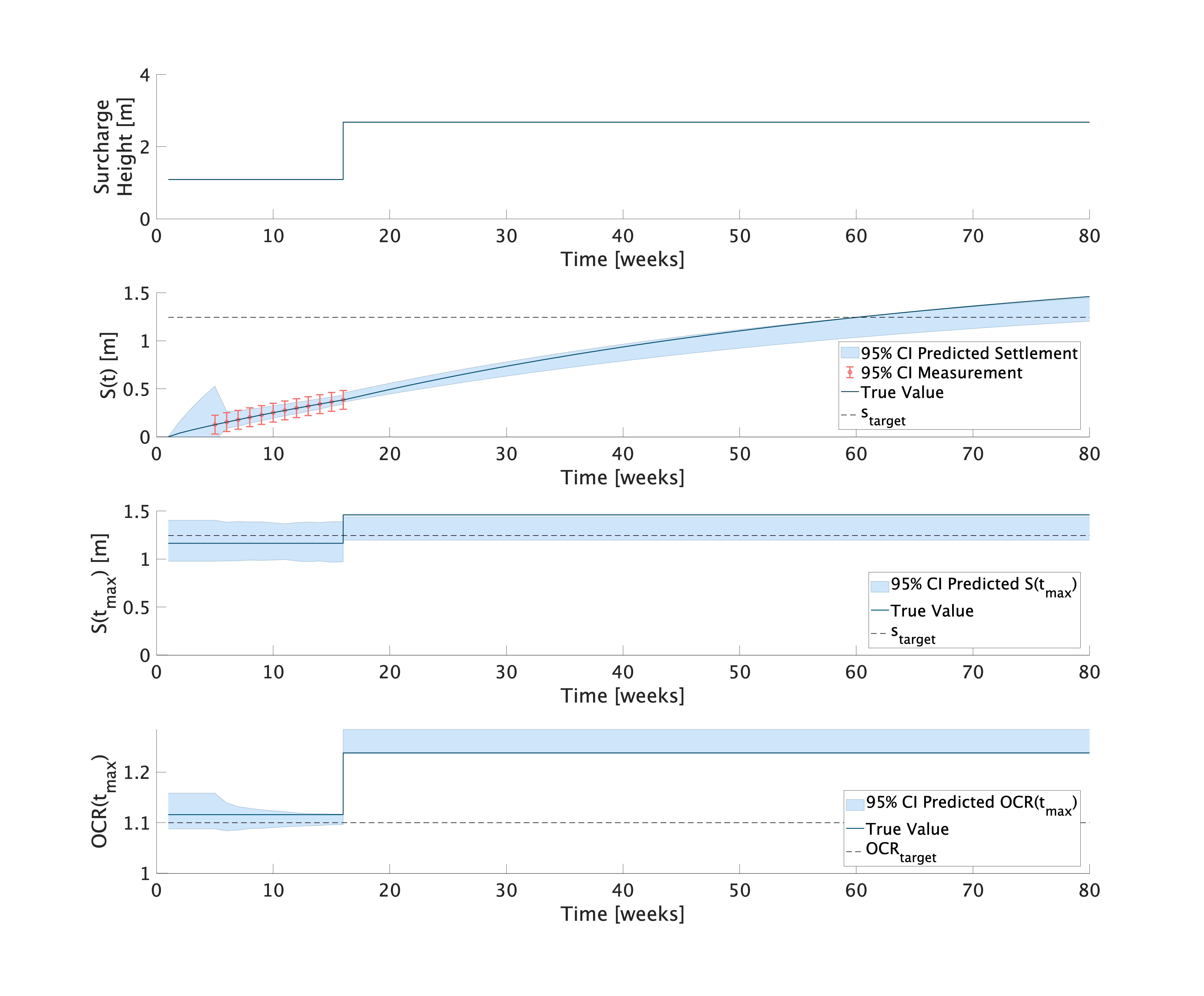}
\caption{The PDT dashboard for an example trajectory with the following heuristic parameters $\mathbf{w}=\left[h_0=1.09\,m, \text{cov}_{th}=0.05, \text{P}_{th}=0.43 \right]$}
\label{fig:heuristic_example}
\end{figure*}

An initial surcharge height $a_0$ is established at the start of the project. Subsequently, the settlement induced by the surcharge is monitored at weekly intervals, denoted as $t$. Following each observation, Bayesian updating of the physical state $\mathbf{X}_t$ is performed as detailed in \Cref{sec:bayesian_learning}. Data is collected until the standard deviation of the updated belief falls below a predefined threshold $\text{cov}_{th}$. Utilizing the posterior belief of the settlement $S(t_{max})$, the probability of not achieving $s_{target}$ by the time $t_{max}$ is assessed. If this probability is higher than a probability threshold $\text{P}_{th}$, the surcharge height is increased by $h_{sur}$, which is defined as the smallest value that ensures that the probability $\Pr\left[S(t_{max})>s_{target}\right]$ falls below this threshold. 

An example outcome of the PDT for the described scenario is depicted in \Cref{fig:heuristic_example}. The first panel shows the evolution of the surcharge height over time $t$. The second panel outlines the belief state evolution of the settlement as observations are collected. Measurements continue until $t=16$ when the confidence interval (CI) for settlement predictions falls below the threshold $\text{cov}_{th}$. This indicates that prediction uncertainty has been sufficiently minimized for decision-making. Consequently, the surcharge height is increased to ensure that the predicted probability of not reaching $s_{target}$ is lower than $\text{P}_{th}$. The third and fourth panels depict the evolution of the $95\%$ CI together with the true value for the quantities of interest $S(t_{max})$ and $OCR(t_{max})$. They demonstrate that the mitigation action was effective, as $s_{target}$ is now reached within the designated project time.

To identify the optimal heuristic parameters $\mathbf{w}=\left[h_0, \text{cov}_{th}, \text{P}_{th}\right]$ that minimize the expected total cost introduced in \Cref{eq:opt_heuristic} a CE optimization is performed.

\subsection{Computational details}

For the CE method, specific values are set as follows: number of CE samples per iteration $n_{CE}=10^2$, maximum number of iterations $n_E=50$, number of MCS samples for each CE set $n_{MC}=10^2$ and number of samples considered for Bayesian updating $n_{BU}=10^2$. The expected cost associated with the optimized heuristic parameters is assessed in a more extensive evaluation involving $n_{MC}=5*10^3$ Monte Carlo samples. The optimization is conducted using a computing setup comprising a ten-core 3.2 GHz CPU and a sixteen-core GPU, equipped with 32GB of memory, resulting in a total computation time of 21 minutes. This offline computation was executed once per optimization run.

\section{Results}
\label{sec:results}
\subsection{Varying measurement error}

\begin{table}[h!]
\caption{Optimal heuristic parameters and associated expected costs for the three scenarios with varying measurement errors investigated in this work.}
\label{tab:results_scenarios}
\centering
\begin{adjustbox}{width=0.8\textwidth}
\begin{tabular}{l l | c c c}
Parameter&Unit & $\sigma_\epsilon=0.05\text{\,m}$ &
$\sigma_\epsilon=0.10\text{\,m}$ & $\sigma_\epsilon=0.15\text{\,m}$\\
\hline
$h_0$&[m]& $0.98$ & $0.99$ & $1.06$\\
$COV_{th}$& [-]& $0.50$ & $0.50$ & $0.49$\\
$P_{th}$& [-]& $0.62$ & $0.47$ & $0.40$\\
\hline
Expected cost&[$10^{6} \text{ SEK}$] & $6.42$ & $6.51$ & $6.88$ \\
Std. dev. cost&[$10^{6} \text{ SEK}$] & $5.29$ & $5.38$ & $4.94$ \\
\hline
\end{tabular}
\end{adjustbox}
\end{table}

The proposed heuristic is applied for three scenarios, distinguished by the standard deviation of the measurement error, which varies between $\sigma_\varepsilon=0.05m-0.15m$. \Cref{tab:results_scenarios} presents the resulting expected costs and parameters of the optimized heuristic for each scenario.

\begin{figure}[h]
\centering
\includegraphics[width=0.5\textwidth]{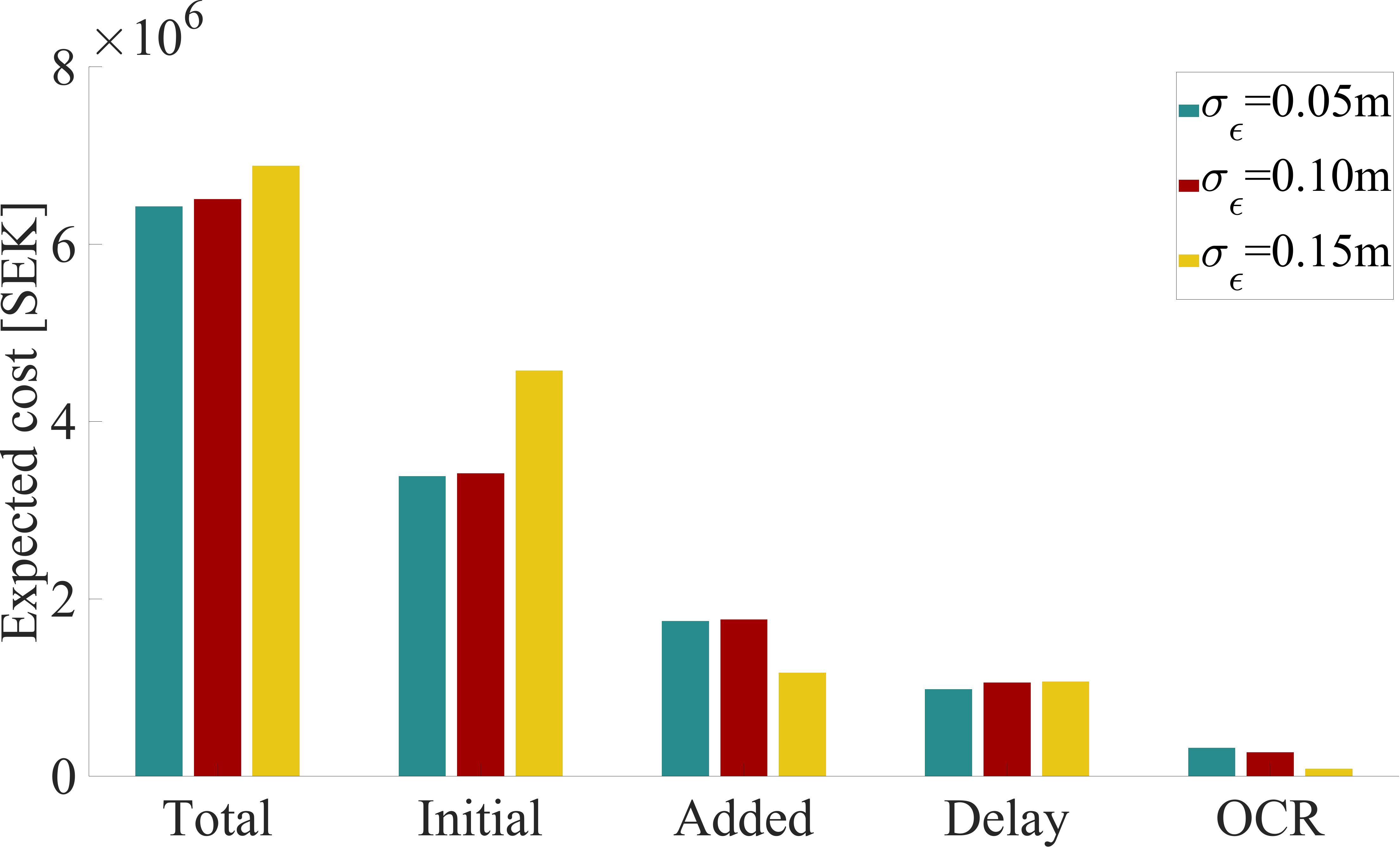}
\caption{Cost breakdown of the best results for each scenario for the proposed heuristic.}
\label{fig:cost_breakdown}
\end{figure}

A side-to-side breakdown of the expected costs for each scenario, provided in \Cref{fig:cost_breakdown}, indicates similar performance for measurement errors $\sigma_\varepsilon = 0.05m$ and $\sigma_\varepsilon = 0.10m$, with a noticeable increase in expected costs for $\sigma_\varepsilon = 0.15m$. However, the latter case results in a higher expected cost, but with a lower standard deviation. This is probably due to the low robustness of the CE method encountered for this problem. Nonetheless, the heuristic consistently extracts actionable information for decision-making across all scenarios.

\begin{figure*}[h]
\centering
\includegraphics[width=0.5\textwidth]{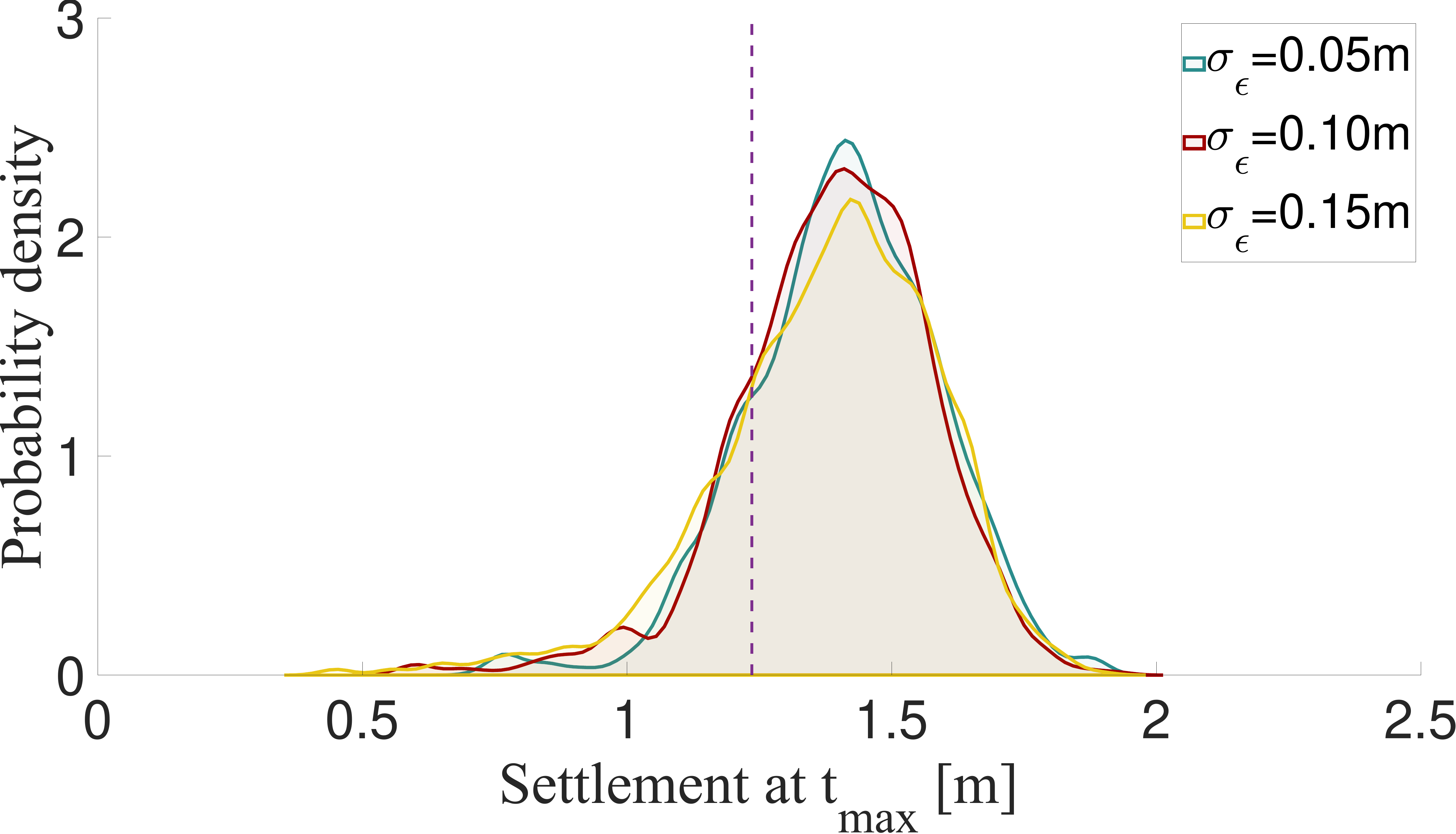}%
\includegraphics[width=0.5\textwidth]{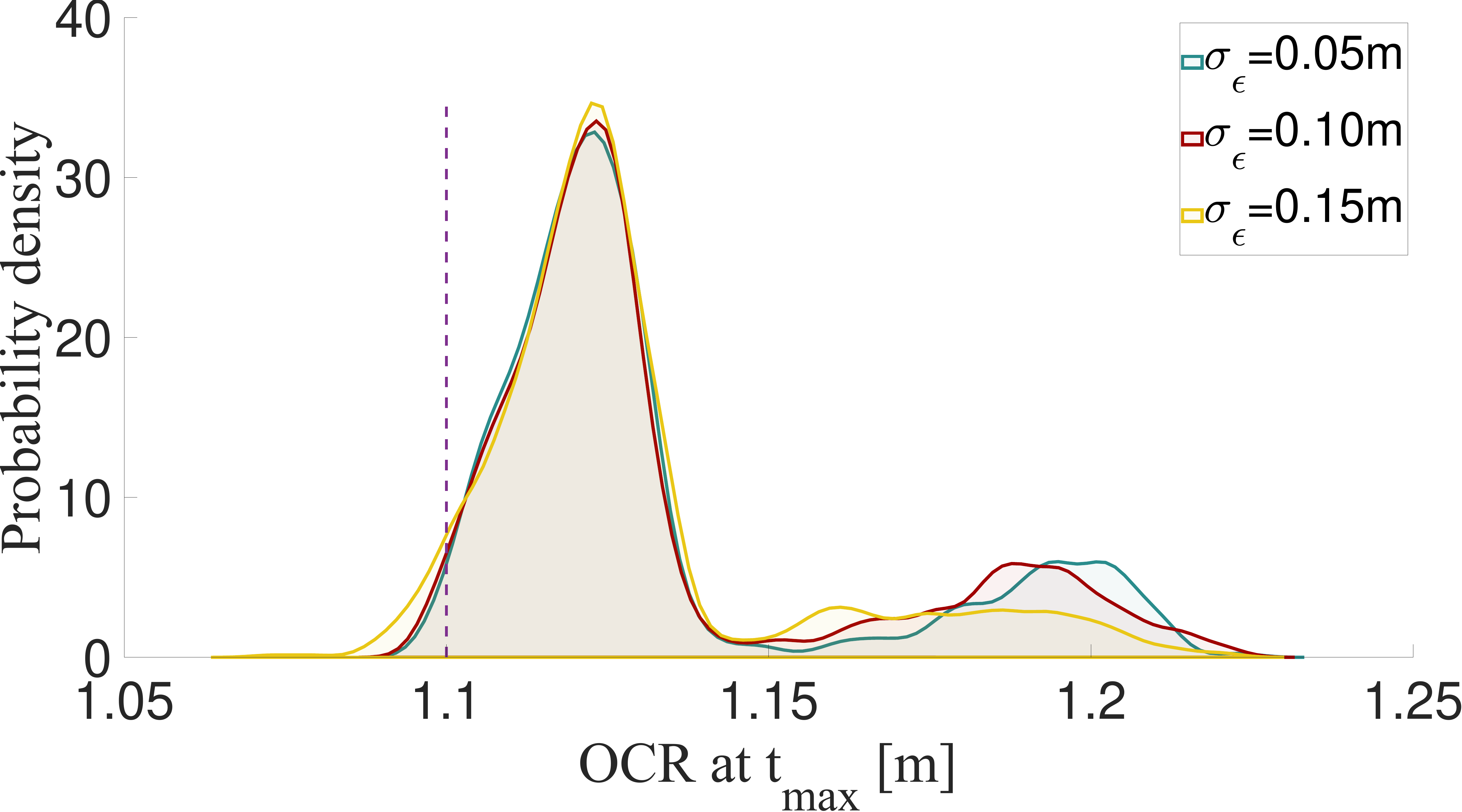}
\caption{Kernel representation of the a) settlement and b) OCR at time of unloading $t_{max}$ for the best results of each scenario for the proposed heuristic. $10\, 0000$ samples where obtained with the parameters listed in \Cref{tab:results}.}
\label{fig:stmax_ocrtmax_example}
\end{figure*}

Furthermore, \Cref{fig:stmax_ocrtmax_example} showcases the distribution of settlement and OCR at the time of unloading across all scenarios. We note that for each scenario the heuristic ensures compliance with both $s_{target}$ and $OCR_{target}$ requirements upon unloading for most cases. The peaks of the probability distributions are close to the requirements, indicating the efficacy of the heuristic in identifying solutions that not only meet the specified criteria but also optimize cost efficiency.

\subsection{Comparison with state-of-the-art}

In previous work, measurement errors were only assumed in \citet{cotoarba2023data}, and even minimal measurement errors significantly impacted the efficiency of the employed heuristics because they were based on observations directly instead of predictive modeling. These heuristics worked well for \textit{ideal} conditions, where measurement errors were omitted.

The comparison summarized in \Cref{tab:results} shows the results for the current heuristic against the results \textit{Heuristic 1} and \textit{Heuristic 4} from previous studies by \citet{bismut2023optimal,cotoarba2023data}. \textit{Heuristic 1} serves as the reference case and offers a simplistic decision framework that optimizes only the initial surcharge height without the possibility for subsequent adjustments. This can be considered the state-of-the-art, as a Monte Carlo Simulation is used for optimization. \textit{Heuristic 4} adopts a regression model, trained on available measurements at the decision time to predict $S_{t_{max}}$. The surcharge height is adjusted according to the difference between the prediction and $s_{target}$. For \textit{Heuristic 4} the largest measurement error that was considered is $\sigma_\varepsilon=0.03 \, m$.

\begin{table}[h]
\caption{Comparison between best results for Probabilistic-Digital-Twin-based heuristics and two heuristics from previous work.}
\label{tab:results}
\centering
\begin{adjustbox}{width=0.8\textwidth}
\begin{tabular}{l l | c c c | c c}
& & \multicolumn{3}{c|}{Probabilistic} & \multicolumn{2}{c}{State-of-the-art}\\
& & \multicolumn{3}{c|}{Digital Twin} & \multicolumn{2}{c}{\citep{cotoarba2023data}}\\
\hline
Parameter&Unit& \multicolumn{3}{c|}{Heuristic BU} & Heuristic 1 & Heuristic 4\\
\hline
$\sigma_\epsilon$& [m] & $0.05$ & $0.10$ & $0.15$ & - & $0.03$\\
\hline
Expected cost&[$10^{6} \text{ SEK}$] & $6.42$ & $6.51$ & $6.88$ & $8.11$ & $7.33$\\
Std. dev. cost&[$10^{6} \text{ SEK}$] & $5.29$ & $5.38$ & $4.94$ & $7.40$ & $5.56$\\
\hline
\end{tabular}
\end{adjustbox}
\end{table}

In comparison with previous work, even the worst-case scenario for the PDT-based heuristic with $\sigma_\varepsilon=0.15m$ outperforms Heuristic 4 with $\sigma_\varepsilon=0.03m$, achieving a reduction in the expected total costs by $6\text{-}13\%$ and up to $20\%$ compared to the state-of-the-art. Additionally, the standard deviation of the cost is reduced by up to $40\%$ and shows that by incorporating Bayesian learning, variability in the final expected cost can be reduced.

\section{Discussion}
\label{sec:discussion}
This example application of a PDT demonstrates the potential of applying Bayesian learning to optimize geotechnical construction under uncertainty and sequential data input. The approach effectively extracts information from noisy data, enables more accurate predictions of settlements and OCR, and identifies cost-efficient construction strategies using a heuristic strategy optimization.

However, while promising, this first implementation of the PDT has several limitations. First, property data is only used for initial model development due to limited availability. The model can be easily extended to update the digital state as new property data (e.g., soil samples) becomes available. Second,  extending the subsoil model from 1D to 2D or 3D would enable more accurate representation of soil conditions. 
Third, the implementation currently considers only one behavioral prediction model (soil settlement over time under load), whereas the full potential of the PDT approach lies in integrating multiple models. Increasing PDT complexity in this way will require the investigation of more efficient data models and methods. 

Despite these limitations and challenges, this work demonstrates the potential of the PDT approach to enhance geotechnical design and construction decision-making compared to traditional methods. The PDT integrates both traditional and novel geotechnical design and construction methods. The adoption of stochastic parameter description, enables the integration of Bayesian probabilistic models and learning from diverse data sources. This aligns with established geotechnical design practices, such as the observational method outlined in Eurocode 7 \citep{CEN_04}. In addition to model updating, the probabilistic description allows the use of existing optimization approaches for sequential decision-making under uncertainties.

\section{Conclusion}
\label{sec:conclusion}

Findings from the application of the probabilistic digital twin to an embankment problem highlight its potential for geotechnical design and construction. The PDT framework demonstrates compatibility with prevalent design practices within geotechnical engineering (e.g., observational method) and offers a structured approach for explicitly managing uncertainties and effectively customizing models to reflect site-specific conditions to improve decision-making.

The PDT framework emerges as a natural extension of the traditional DT approach to offer a robust mechanism for the AECOM industries to navigate the complexities of project-specific uncertainties. It shows its largest benefits when scaled and designed to be generally applicable in AECOM industries. It can be used to optimize design and construction processes and enhance decision-making during the life-cycle of a system, marking a significant step forward in the evolution of digital twin technologies.

\paragraph{Data availability statement}
Data availability is not applicable to this article as no new data were created or analyzed in this study.

\paragraph{Competing interest} The authors declare no competing interests exist.

\paragraph{Funding Statement}
This research is supported by the TUM Georg Nemetschek Institute Artificial Intelligence for the Built World.


\bibliographystyle{elsarticle-harv}
\bibliography{bibliography.bib}

\begin{thebibliography}{66}
\expandafter\ifx\csname natexlab\endcsname\relax\def\natexlab#1{#1}\fi
\providecommand{\url}[1]{\texttt{#1}}
\providecommand{\href}[2]{#2}
\providecommand{\path}[1]{#1}
\providecommand{\DOIprefix}{doi:}
\providecommand{\ArXivprefix}{arXiv:}
\providecommand{\URLprefix}{URL: }
\providecommand{\Pubmedprefix}{pmid:}
\providecommand{\doi}[1]{\href{http://dx.doi.org/#1}{\path{#1}}}
\providecommand{\Pubmed}[1]{\href{pmid:#1}{\path{#1}}}
\providecommand{\bibinfo}[2]{#2}
\ifx\xfnm\relax \def\xfnm[#1]{\unskip,\space#1}\fi
\bibitem[{Abualdenien and Borrmann(2022)}]{abualdenien2022levels}
\bibinfo{author}{Abualdenien, J.}, \bibinfo{author}{Borrmann, A.},
  \bibinfo{year}{2022}.
\newblock \bibinfo{title}{Levels of detail, development, definition, and
  information need: A critical literature review.}
\newblock \bibinfo{journal}{Journal of Information Technology in Construction}
  \bibinfo{volume}{27}.
\newblock \DOIprefix\doi{https://dx.doi.org/10.36680/j.itcon.2022.018}.
\bibitem[{Agrell et~al.(2023)Agrell, Rognlien~Dahl and
  Hafver}]{agrell2023optimal}
\bibinfo{author}{Agrell, C.}, \bibinfo{author}{Rognlien~Dahl, K.},
  \bibinfo{author}{Hafver, A.}, \bibinfo{year}{2023}.
\newblock \bibinfo{title}{Optimal sequential decision making with probabilistic
  digital twins: Theoretical foundations}.
\newblock \bibinfo{journal}{SN Applied Sciences} \bibinfo{volume}{5},
  \bibinfo{pages}{114}.
\newblock \DOIprefix\doi{https://doi.org/10.1007/s42452-023-05316-9}.
\bibitem[{Alibrandi(2022)}]{alibrandi2022risk}
\bibinfo{author}{Alibrandi, U.}, \bibinfo{year}{2022}.
\newblock \bibinfo{title}{Risk-informed digital twin of buildings and
  infrastructures for sustainable and resilient urban communities}.
\newblock \bibinfo{journal}{ASCE-ASME Journal of Risk and Uncertainty in
  Engineering Systems, Part A: Civil Engineering} \bibinfo{volume}{8},
  \bibinfo{pages}{04022032}.
\newblock \DOIprefix\doi{https://doi.org/10.1061/AJRUA6.0001238}.
\bibitem[{Andriotis and Papakonstantinou(2019)}]{andriotis2019managing}
\bibinfo{author}{Andriotis, C.P.}, \bibinfo{author}{Papakonstantinou, K.G.},
  \bibinfo{year}{2019}.
\newblock \bibinfo{title}{Managing engineering systems with large state and
  action spaces through deep reinforcement learning}.
\newblock \bibinfo{journal}{Reliability Engineering \& System Safety}
  \bibinfo{volume}{191}, \bibinfo{pages}{106483}.
\bibitem[{Benjamin and Cornell(2014)}]{benjamin2014probability}
\bibinfo{author}{Benjamin, J.R.}, \bibinfo{author}{Cornell, C.A.},
  \bibinfo{year}{2014}.
\newblock \bibinfo{title}{Probability, statistics, and decision for civil
  engineers}.
\newblock \bibinfo{publisher}{Courier Corporation}.
\bibitem[{Bismut et~al.(2023)Bismut, Cotoarb{\u{a}}, Spross and
  Straub}]{bismut2023optimal}
\bibinfo{author}{Bismut, E.}, \bibinfo{author}{Cotoarb{\u{a}}, D.},
  \bibinfo{author}{Spross, J.}, \bibinfo{author}{Straub, D.},
  \bibinfo{year}{2023}.
\newblock \bibinfo{title}{Optimal adaptive decision rules in geotechnical
  construction considering uncertainty}.
\newblock \bibinfo{journal}{G{\'e}otechnique} , \bibinfo{pages}{1--12}.
\bibitem[{Bismut and Straub(2021)}]{bismut2021optimal}
\bibinfo{author}{Bismut, E.}, \bibinfo{author}{Straub, D.},
  \bibinfo{year}{2021}.
\newblock \bibinfo{title}{Optimal adaptive inspection and maintenance planning
  for deteriorating structural systems}.
\newblock \bibinfo{journal}{Reliability Engineering \& System Safety}
  \bibinfo{volume}{215}, \bibinfo{pages}{107891}.
\bibitem[{Bismut et~al.(2022)Bismut, Straub and Pandey}]{Bismut_et_al_22}
\bibinfo{author}{Bismut, E.}, \bibinfo{author}{Straub, D.},
  \bibinfo{author}{Pandey, M.}, \bibinfo{year}{2022}.
\newblock \bibinfo{title}{Inspection and maintenance planning of a feeder
  piping system}.
\newblock \bibinfo{journal}{Reliability Engineering \& System Safety}
  \bibinfo{volume}{224}, \bibinfo{pages}{108521}.
\bibitem[{Brilakis et~al.(2019)Brilakis, Pan, Borrmann, Mayer, Rhein, Vos,
  Pettinato and Wagner}]{brilakisborrman2019}
\bibinfo{author}{Brilakis, I.}, \bibinfo{author}{Pan, Y.},
  \bibinfo{author}{Borrmann, A.}, \bibinfo{author}{Mayer, H.G.},
  \bibinfo{author}{Rhein, F.}, \bibinfo{author}{Vos, C.},
  \bibinfo{author}{Pettinato, E.}, \bibinfo{author}{Wagner, S.},
  \bibinfo{year}{2019}.
\newblock \bibinfo{title}{Built Environment Digital Twining}.
\newblock \URLprefix \url{https://www.repository.cam.ac.uk/handle/1810/318329},
  \DOIprefix\doi{10.17863/CAM.65445}.
\bibitem[{Capp{\'e} et~al.(2007)Capp{\'e}, Godsill and
  Moulines}]{cappe2007overview}
\bibinfo{author}{Capp{\'e}, O.}, \bibinfo{author}{Godsill, S.J.},
  \bibinfo{author}{Moulines, E.}, \bibinfo{year}{2007}.
\newblock \bibinfo{title}{An overview of existing methods and recent advances
  in sequential monte carlo}.
\newblock \bibinfo{journal}{Proceedings of the IEEE} \bibinfo{volume}{95},
  \bibinfo{pages}{899--924}.
\bibitem[{CEN EN 1997-1:2004()}]{CEN_04}
CEN EN 1997-1:2004, \bibinfo{year}{2004}.
\newblock \bibinfo{title}{{Eurocode 7: Geotechnical design – Part 1: General
  rules}}.
\newblock \bibinfo{type}{{}}. European Committee for Standardisation.
\bibitem[{Chandler(2011)}]{Chandler2011}
\bibinfo{author}{Chandler, R.J.}, \bibinfo{year}{2011}.
\newblock \bibinfo{title}{Geotechnical data transfer and management for large
  construction projects and national archives}, in:
  \bibinfo{booktitle}{International Symposium on Advances in Ground Technology
  Geo-Information (IS-AGTG 2011)}.
\bibitem[{Chaudhuri et~al.(2023)Chaudhuri, Pash, Hormuth, Lorenzo, Kapteyn, Wu,
  Lima, Yankeelov, Willcox et~al.}]{chaudhuri2023predictive}
\bibinfo{author}{Chaudhuri, A.}, \bibinfo{author}{Pash, G.},
  \bibinfo{author}{Hormuth, D.A.}, \bibinfo{author}{Lorenzo, G.},
  \bibinfo{author}{Kapteyn, M.}, \bibinfo{author}{Wu, C.},
  \bibinfo{author}{Lima, E.A.}, \bibinfo{author}{Yankeelov, T.E.},
  \bibinfo{author}{Willcox, K.}, et~al., \bibinfo{year}{2023}.
\newblock \bibinfo{title}{Predictive digital twin for optimizing
  patient-specific radiotherapy regimens under uncertainty in high-grade
  gliomas}.
\newblock \bibinfo{journal}{Frontiers in Artificial Intelligence}
  \bibinfo{volume}{6}.
\bibitem[{Chopin et~al.(2020)Chopin, Papaspiliopoulos
  et~al.}]{chopin2020introduction}
\bibinfo{author}{Chopin, N.}, \bibinfo{author}{Papaspiliopoulos, O.}, et~al.,
  \bibinfo{year}{2020}.
\newblock \bibinfo{title}{An introduction to sequential Monte Carlo}.
  volume~\bibinfo{volume}{4}.
\newblock \bibinfo{publisher}{Springer}.
\bibitem[{Cotoarb{\u{a}} et~al.(2023)Cotoarb{\u{a}}, Bismut, Spross and
  Straub}]{cotoarba2023data}
\bibinfo{author}{Cotoarb{\u{a}}, D.}, \bibinfo{author}{Bismut, E.},
  \bibinfo{author}{Spross, J.}, \bibinfo{author}{Straub, D.},
  \bibinfo{year}{2023}.
\newblock \bibinfo{title}{Data-driven uncertainty reduction in geotechnical
  engineering: Optimal preloading of a road embankment}, in:
  \bibinfo{booktitle}{14th International Conference on Applications of
  Statistics and Probability in Civil Engineering (ICASP14)}.
\bibitem[{Doucet et~al.(2001a)Doucet, De~Freitas and
  Gordon}]{doucet2001introduction}
\bibinfo{author}{Doucet, A.}, \bibinfo{author}{De~Freitas, N.},
  \bibinfo{author}{Gordon, N.}, \bibinfo{year}{2001}a.
\newblock \bibinfo{title}{An introduction to sequential monte carlo methods}.
\newblock \bibinfo{journal}{Sequential Monte Carlo methods in practice} ,
  \bibinfo{pages}{3--14}.
\bibitem[{Doucet et~al.(2001b)Doucet, De~Freitas, Gordon
  et~al.}]{doucet2001sequential}
\bibinfo{author}{Doucet, A.}, \bibinfo{author}{De~Freitas, N.},
  \bibinfo{author}{Gordon, N.J.}, et~al., \bibinfo{year}{2001}b.
\newblock \bibinfo{title}{Sequential Monte Carlo methods in practice}.
  volume~\bibinfo{volume}{1}.
\newblock \bibinfo{publisher}{Springer}.
\bibitem[{Doucet et~al.(2009)Doucet, Johansen et~al.}]{doucet2009tutorial}
\bibinfo{author}{Doucet, A.}, \bibinfo{author}{Johansen, A.M.}, et~al.,
  \bibinfo{year}{2009}.
\newblock \bibinfo{title}{A tutorial on particle filtering and smoothing:
  Fifteen years later}.
\newblock \bibinfo{journal}{Handbook of nonlinear filtering}
  \bibinfo{volume}{12}, \bibinfo{pages}{3}.
\bibitem[{Gong et~al.(2020)Gong, Zhao, Juang, Tang, Wang and Hu}]{Gong2020}
\bibinfo{author}{Gong, W.}, \bibinfo{author}{Zhao, C.}, \bibinfo{author}{Juang,
  C.H.}, \bibinfo{author}{Tang, H.}, \bibinfo{author}{Wang, H.},
  \bibinfo{author}{Hu, X.}, \bibinfo{year}{2020}.
\newblock \bibinfo{title}{Stratigraphic uncertainty modelling with random field
  approach}.
\newblock \bibinfo{journal}{Computers and Geotechnics} \bibinfo{volume}{125}.
\newblock \DOIprefix\doi{10.1016/j.compgeo.2020.103681}.
\bibitem[{Gramacy(2020)}]{gramacy2020surrogates}
\bibinfo{author}{Gramacy, R.B.}, \bibinfo{year}{2020}.
\newblock \bibinfo{title}{Surrogates: Gaussian process modeling, design, and
  optimization for the applied sciences}.
\newblock \bibinfo{publisher}{Chapman and Hall/CRC}.
\bibitem[{Grieves(2002)}]{grieves2002plm}
\bibinfo{author}{Grieves, M.W.}, \bibinfo{year}{2002}.
\newblock \bibinfo{title}{Plm initiatives}.
\newblock \bibinfo{howpublished}{PowerPoint Slides}.
\newblock \bibinfo{note}{Product Lifecycle Management Special Meeting,
  University of Michigan Lurie Engineering Center}.
\bibitem[{Jensen and Nielsen(2007)}]{jensen2007bayesian}
\bibinfo{author}{Jensen, F.V.}, \bibinfo{author}{Nielsen, T.D.},
  \bibinfo{year}{2007}.
\newblock \bibinfo{title}{Bayesian networks and decision graphs}.
  volume~\bibinfo{volume}{2}.
\newblock \bibinfo{publisher}{Springer}.
\bibitem[{Kamariotis et~al.(2023)Kamariotis, Sardi, Papaioannou, Chatzi and
  Straub}]{kamariotis2023off}
\bibinfo{author}{Kamariotis, A.}, \bibinfo{author}{Sardi, L.},
  \bibinfo{author}{Papaioannou, I.}, \bibinfo{author}{Chatzi, E.},
  \bibinfo{author}{Straub, D.}, \bibinfo{year}{2023}.
\newblock \bibinfo{title}{On off-line and on-line bayesian filtering for
  uncertainty quantification of structural deterioration}.
\newblock \bibinfo{journal}{Data-Centric Engineering} \bibinfo{volume}{4},
  \bibinfo{pages}{e17}.
\bibitem[{Kapteyn et~al.(2021)Kapteyn, Pretorius and
  Willcox}]{kapteyn2021probabilistic}
\bibinfo{author}{Kapteyn, M.G.}, \bibinfo{author}{Pretorius, J.V.},
  \bibinfo{author}{Willcox, K.E.}, \bibinfo{year}{2021}.
\newblock \bibinfo{title}{A probabilistic graphical model foundation for
  enabling predictive digital twins at scale}.
\newblock \bibinfo{journal}{Nature Computational Science} \bibinfo{volume}{1},
  \bibinfo{pages}{337--347}.
\bibitem[{Karniadakis et~al.(2021)Karniadakis, Kevrekidis, Lu, Perdikaris, Wang
  and Yang}]{karniadakis2021physics}
\bibinfo{author}{Karniadakis, G.E.}, \bibinfo{author}{Kevrekidis, I.G.},
  \bibinfo{author}{Lu, L.}, \bibinfo{author}{Perdikaris, P.},
  \bibinfo{author}{Wang, S.}, \bibinfo{author}{Yang, L.}, \bibinfo{year}{2021}.
\newblock \bibinfo{title}{Physics-informed machine learning}.
\newblock \bibinfo{journal}{Nature Reviews Physics} \bibinfo{volume}{3},
  \bibinfo{pages}{422--440}.
\bibitem[{Kitagawa(1998)}]{kitagawa1998self}
\bibinfo{author}{Kitagawa, G.}, \bibinfo{year}{1998}.
\newblock \bibinfo{title}{A self-organizing state-space model}.
\newblock \bibinfo{journal}{Journal of the American Statistical Association} ,
  \bibinfo{pages}{1203--1215}.
\bibitem[{Kochenderfer(2015)}]{kochenderfer2015decision}
\bibinfo{author}{Kochenderfer, M.J.}, \bibinfo{year}{2015}.
\newblock \bibinfo{title}{Decision making under uncertainty: theory and
  application}.
\newblock \bibinfo{publisher}{MIT press}.
\bibitem[{Kochunas and Huan(2021)}]{kochunas2021digital}
\bibinfo{author}{Kochunas, B.}, \bibinfo{author}{Huan, X.},
  \bibinfo{year}{2021}.
\newblock \bibinfo{title}{Digital twin concepts with uncertainty for nuclear
  power applications}.
\newblock \bibinfo{journal}{Energies} \bibinfo{volume}{14},
  \bibinfo{pages}{4235}.
\bibitem[{Koller and Friedman(2009)}]{koller2009probabilistic}
\bibinfo{author}{Koller, D.}, \bibinfo{author}{Friedman, N.},
  \bibinfo{year}{2009}.
\newblock \bibinfo{title}{Probabilistic graphical models: principles and
  techniques}.
\newblock \bibinfo{publisher}{MIT press}.
\bibitem[{Kritzinger et~al.(2018)Kritzinger, Karner, Traar, Henjes and
  Sihn}]{Kritzinger2018}
\bibinfo{author}{Kritzinger, W.}, \bibinfo{author}{Karner, M.},
  \bibinfo{author}{Traar, G.}, \bibinfo{author}{Henjes, J.},
  \bibinfo{author}{Sihn, W.}, \bibinfo{year}{2018}.
\newblock \bibinfo{title}{Digital twin in manufacturing: A categorical
  literature review and classification}.
\newblock \bibinfo{journal}{Ifac-PapersOnline} \bibinfo{volume}{51},
  \bibinfo{pages}{1016--1022}.
\bibitem[{Kroese et~al.(2006)Kroese, Porotsky and Rubinstein}]{Kroese_et_al_06}
\bibinfo{author}{Kroese, D.P.}, \bibinfo{author}{Porotsky, S.},
  \bibinfo{author}{Rubinstein, R.Y.}, \bibinfo{year}{2006}.
\newblock \bibinfo{title}{The cross-entropy method for continuous
  multi-extremal optimization}.
\newblock \bibinfo{journal}{Methodology and Computing in Applied Probability}
  \bibinfo{volume}{8}, \bibinfo{pages}{383--407}.
\bibitem[{Kulhawy et~al.(2006)Kulhawy, Phoon, Prakoso and Hirany}]{Kulhawy2006}
\bibinfo{author}{Kulhawy, F.H.}, \bibinfo{author}{Phoon, K.K.},
  \bibinfo{author}{Prakoso, W.A.}, \bibinfo{author}{Hirany, A.},
  \bibinfo{year}{2006}.
\newblock \bibinfo{title}{Reliability-based design of foundations for
  transmission line structures}.
\newblock \bibinfo{journal}{Electrical Transmission Line and Substation
  Structures: Structural Reliability in a Changing World - Proceedings of the
  2006 Electrical Transmission Conference} \bibinfo{volume}{218},
  \bibinfo{pages}{184--194}.
\newblock \DOIprefix\doi{10.1061/40790(218)17}.
\bibitem[{Li et~al.(2015)Li, Li, Ji and Dai}]{li2015kalman}
\bibinfo{author}{Li, Q.}, \bibinfo{author}{Li, R.}, \bibinfo{author}{Ji, K.},
  \bibinfo{author}{Dai, W.}, \bibinfo{year}{2015}.
\newblock \bibinfo{title}{Kalman filter and its application}, in:
  \bibinfo{booktitle}{2015 8th international conference on intelligent networks
  and intelligent systems (ICINIS)}, \bibinfo{organization}{IEEE}. pp.
  \bibinfo{pages}{74--77}.
\bibitem[{Liu et~al.(2021)Liu, Fang, Dong and Xu}]{liu2021review}
\bibinfo{author}{Liu, M.}, \bibinfo{author}{Fang, S.}, \bibinfo{author}{Dong,
  H.}, \bibinfo{author}{Xu, C.}, \bibinfo{year}{2021}.
\newblock \bibinfo{title}{Review of digital twin about concepts, technologies,
  and industrial applications}.
\newblock \bibinfo{journal}{Journal of Manufacturing Systems}
  \bibinfo{volume}{58}, \bibinfo{pages}{346--361}.
\bibitem[{Nath and Mahadevan(2022)}]{nath2022probabilistic}
\bibinfo{author}{Nath, P.}, \bibinfo{author}{Mahadevan, S.},
  \bibinfo{year}{2022}.
\newblock \bibinfo{title}{Probabilistic digital twin for additive manufacturing
  process design and control}.
\newblock \bibinfo{journal}{Journal of Mechanical Design}
  \bibinfo{volume}{144}, \bibinfo{pages}{091704}.
\bibitem[{Norris(1998)}]{norris1998markov}
\bibinfo{author}{Norris, J.R.}, \bibinfo{year}{1998}.
\newblock \bibinfo{title}{Markov chains}.
\newblock \bibinfo{number}{2}, \bibinfo{publisher}{Cambridge university press}.
\bibitem[{Opoku et~al.(2021)Opoku, Perera, Osei-Kyei and
  Rashidi}]{opoku2021digital}
\bibinfo{author}{Opoku, D.G.J.}, \bibinfo{author}{Perera, S.},
  \bibinfo{author}{Osei-Kyei, R.}, \bibinfo{author}{Rashidi, M.},
  \bibinfo{year}{2021}.
\newblock \bibinfo{title}{Digital twin application in the construction
  industry: A literature review}.
\newblock \bibinfo{journal}{Journal of Building Engineering}
  \bibinfo{volume}{40}, \bibinfo{pages}{102726}.
\bibitem[{Papakonstantinou et~al.(2018)Papakonstantinou, Andriotis and
  Shinozuka}]{papakonstantinou2018pomdp}
\bibinfo{author}{Papakonstantinou, K.G.}, \bibinfo{author}{Andriotis, C.P.},
  \bibinfo{author}{Shinozuka, M.}, \bibinfo{year}{2018}.
\newblock \bibinfo{title}{Pomdp and momdp solutions for structural life-cycle
  cost minimization under partial and mixed observability}.
\newblock \bibinfo{journal}{Structure and Infrastructure Engineering}
  \bibinfo{volume}{14}, \bibinfo{pages}{869--882}.
\bibitem[{Peck(1969)}]{peck1969advantages}
\bibinfo{author}{Peck, R.B.}, \bibinfo{year}{1969}.
\newblock \bibinfo{title}{Advantages and limitations of the observational
  method in applied soil mechanics}.
\newblock \bibinfo{journal}{Geotechnique} \bibinfo{volume}{19},
  \bibinfo{pages}{171--187}.
\bibitem[{Phoon()}]{Phoon2019}
\bibinfo{author}{Phoon, K.K.}, .
\newblock \bibinfo{title}{The story of statistics in geotechnical engineering}.
\newblock \bibinfo{journal}{Georisk: Assessment and Management of Risk for
  Engineered Systems and Geohazards} ,
  \bibinfo{pages}{3--25}\DOIprefix\doi{10.1080/17499518.2019.1700423}.
\bibitem[{Phoon et~al.(2022a)Phoon, Cao, Ji, Leung, Najjar, Shuku, Tang, Yin,
  Ikumasa and Ching}]{phoon2022geotechnical}
\bibinfo{author}{Phoon, K.K.}, \bibinfo{author}{Cao, Z.J.},
  \bibinfo{author}{Ji, J.}, \bibinfo{author}{Leung, Y.F.},
  \bibinfo{author}{Najjar, S.}, \bibinfo{author}{Shuku, T.},
  \bibinfo{author}{Tang, C.}, \bibinfo{author}{Yin, Z.Y.},
  \bibinfo{author}{Ikumasa, Y.}, \bibinfo{author}{Ching, J.},
  \bibinfo{year}{2022}a.
\newblock \bibinfo{title}{Geotechnical uncertainty, modeling, and decision
  making}.
\newblock \bibinfo{journal}{Soils and Foundations} \bibinfo{volume}{62},
  \bibinfo{pages}{101189}.
\bibitem[{Phoon et~al.(2022b)Phoon, Ching and Cao}]{phoon2022unpacking}
\bibinfo{author}{Phoon, K.K.}, \bibinfo{author}{Ching, J.},
  \bibinfo{author}{Cao, Z.}, \bibinfo{year}{2022}b.
\newblock \bibinfo{title}{Unpacking data-centric geotechnics}.
\newblock \bibinfo{journal}{Underground space} \bibinfo{volume}{7},
  \bibinfo{pages}{967--989}.
\bibitem[{Phoon and Kulhawy(1999)}]{Phoon1999}
\bibinfo{author}{Phoon, K.K.}, \bibinfo{author}{Kulhawy, F.H.},
  \bibinfo{year}{1999}.
\newblock \bibinfo{title}{Characterization of geotechnical variability}.
\newblock \bibinfo{journal}{Canadian geotechnical journal}
  \bibinfo{volume}{36}, \bibinfo{pages}{612--624}.
\bibitem[{Porta et~al.(2005)Porta, Spaan and Vlassis}]{porta2005robot}
\bibinfo{author}{Porta, J.M.}, \bibinfo{author}{Spaan, M.T.},
  \bibinfo{author}{Vlassis, N.}, \bibinfo{year}{2005}.
\newblock \bibinfo{title}{Robot planning in partially observable continuous
  domains}.
\newblock \bibinfo{journal}{1st Robotics: Science and Systems Conference} ,
  \bibinfo{pages}{217--224}.
\bibitem[{Rasmussen and Williams(2005)}]{Rasmussen2005}
\bibinfo{author}{Rasmussen, C.E.}, \bibinfo{author}{Williams, C.K.I.},
  \bibinfo{year}{2005}.
\newblock \bibinfo{title}{Gaussian processes for machine learning}.
\newblock \bibinfo{journal}{Gaussian Processes for Machine Learning} \URLprefix
  \url{https://direct.mit.edu/books/book/2320/Gaussian-Processes-for-Machine-Learning},
  \DOIprefix\doi{10.7551/MITPRESS/3206.001.0001}.
\bibitem[{Robertson(2009)}]{Robertson2009}
\bibinfo{author}{Robertson, P.K.}, \bibinfo{year}{2009}.
\newblock \bibinfo{title}{Interpretation of cone penetration tests — a
  unified approach}.
\newblock \bibinfo{journal}{Canadian Geotechnical Journal}
  \bibinfo{volume}{46}, \bibinfo{pages}{1337--1355}.
\newblock \DOIprefix\doi{10.1139/T09-065}.
\bibitem[{Roy et~al.(2005)Roy, Gordon and Thrun}]{roy2005finding}
\bibinfo{author}{Roy, N.}, \bibinfo{author}{Gordon, G.},
  \bibinfo{author}{Thrun, S.}, \bibinfo{year}{2005}.
\newblock \bibinfo{title}{Finding approximate pomdp solutions through belief
  compression}.
\newblock \bibinfo{journal}{Journal of artificial intelligence research}
  \bibinfo{volume}{23}, \bibinfo{pages}{1--40}.
\bibitem[{Russell and Norvig(2016)}]{russell2010artificial}
\bibinfo{author}{Russell, S.J.}, \bibinfo{author}{Norvig, P.},
  \bibinfo{year}{2016}.
\newblock \bibinfo{title}{Artificial intelligence: a modern approach}.
\newblock \bibinfo{publisher}{Pearson}.
\bibitem[{Semeraro et~al.(2021)Semeraro, Lezoche, Panetto and
  Dassisti}]{semeraro2021digital}
\bibinfo{author}{Semeraro, C.}, \bibinfo{author}{Lezoche, M.},
  \bibinfo{author}{Panetto, H.}, \bibinfo{author}{Dassisti, M.},
  \bibinfo{year}{2021}.
\newblock \bibinfo{title}{Digital twin paradigm: A systematic literature
  review}.
\newblock \bibinfo{journal}{Computers in Industry} \bibinfo{volume}{130},
  \bibinfo{pages}{103469}.
\bibitem[{Shachter(1986)}]{shachter1986evaluating}
\bibinfo{author}{Shachter, R.D.}, \bibinfo{year}{1986}.
\newblock \bibinfo{title}{Evaluating influence diagrams}.
\newblock \bibinfo{journal}{Operations research} \bibinfo{volume}{34},
  \bibinfo{pages}{871--882}.
\bibitem[{Silver and Veness(2010)}]{silver2010monte}
\bibinfo{author}{Silver, D.}, \bibinfo{author}{Veness, J.},
  \bibinfo{year}{2010}.
\newblock \bibinfo{title}{Monte-carlo planning in large pomdps}.
\newblock \bibinfo{journal}{Advances in neural information processing systems}
  \bibinfo{volume}{23}.
\bibitem[{Singh et~al.(2022)Singh, Srivastava, Fuenmayor, Kuts, Qiao, Murray
  and Devine}]{singh2022applications}
\bibinfo{author}{Singh, M.}, \bibinfo{author}{Srivastava, R.},
  \bibinfo{author}{Fuenmayor, E.}, \bibinfo{author}{Kuts, V.},
  \bibinfo{author}{Qiao, Y.}, \bibinfo{author}{Murray, N.},
  \bibinfo{author}{Devine, D.}, \bibinfo{year}{2022}.
\newblock \bibinfo{title}{Applications of digital twin across industries: A
  review}.
\newblock \bibinfo{journal}{Applied Sciences} \bibinfo{volume}{12},
  \bibinfo{pages}{5727}.
\bibitem[{Song et~al.(2020)Song, Astroza, Ebrahimian, Moaveni and
  Papadimitriou}]{song2020adaptive}
\bibinfo{author}{Song, M.}, \bibinfo{author}{Astroza, R.},
  \bibinfo{author}{Ebrahimian, H.}, \bibinfo{author}{Moaveni, B.},
  \bibinfo{author}{Papadimitriou, C.}, \bibinfo{year}{2020}.
\newblock \bibinfo{title}{Adaptive kalman filters for nonlinear finite element
  model updating}.
\newblock \bibinfo{journal}{Mechanical Systems and Signal Processing}
  \bibinfo{volume}{143}, \bibinfo{pages}{106837}.
\bibitem[{Spross and Johansson(2017)}]{spross2017observational}
\bibinfo{author}{Spross, J.}, \bibinfo{author}{Johansson, F.},
  \bibinfo{year}{2017}.
\newblock \bibinfo{title}{When is the observational method in geotechnical
  engineering favourable?}
\newblock \bibinfo{journal}{Structural Safety} \bibinfo{volume}{66},
  \bibinfo{pages}{17--26}.
\bibitem[{Spross and Larsson(2021)}]{spross2021probabilistic}
\bibinfo{author}{Spross, J.}, \bibinfo{author}{Larsson, S.},
  \bibinfo{year}{2021}.
\newblock \bibinfo{title}{Probabilistic observational method for design of
  surcharges on vertical drains}.
\newblock \bibinfo{journal}{G{\'e}otechnique} \bibinfo{volume}{71},
  \bibinfo{pages}{226--238}.
\bibitem[{Sudret(2014)}]{sudret2014polynomial}
\bibinfo{author}{Sudret, B.}, \bibinfo{year}{2014}.
\newblock \bibinfo{title}{Polynomial chaos expansions and stochastic finite
  element methods}.
\newblock \bibinfo{journal}{Risk and reliability in geotechnical engineering} ,
  \bibinfo{pages}{265--300}.
\bibitem[{Torzoni et~al.(2024)Torzoni, Tezzele, Mariani, Manzoni and
  Willcox}]{torzoni2024digital}
\bibinfo{author}{Torzoni, M.}, \bibinfo{author}{Tezzele, M.},
  \bibinfo{author}{Mariani, S.}, \bibinfo{author}{Manzoni, A.},
  \bibinfo{author}{Willcox, K.E.}, \bibinfo{year}{2024}.
\newblock \bibinfo{title}{A digital twin framework for civil engineering
  structures}.
\newblock \bibinfo{journal}{Computer Methods in Applied Mechanics and
  Engineering} \bibinfo{volume}{418}, \bibinfo{pages}{116584}.
\bibitem[{Wang et~al.(2016)Wang, Harken, Osorio-Murillo, Zhu and
  Rubin}]{Wang2016}
\bibinfo{author}{Wang, C.H.}, \bibinfo{author}{Harken, B.},
  \bibinfo{author}{Osorio-Murillo, C.A.}, \bibinfo{author}{Zhu, H.H.},
  \bibinfo{author}{Rubin, Y.}, \bibinfo{year}{2016}.
\newblock \bibinfo{title}{Bayesian approach for probabilistic site
  characterization assimilating borehole experiments and cone penetration
  tests}.
\newblock \bibinfo{journal}{Engineering Geology} \bibinfo{volume}{207},
  \bibinfo{pages}{1--13}.
\newblock \DOIprefix\doi{10.1016/j.enggeo.2016.04.002}.
\bibitem[{Wang et~al.(2019)Wang, Wang, Wellmann and Liang}]{wang2019bayesian}
\bibinfo{author}{Wang, H.}, \bibinfo{author}{Wang, X.},
  \bibinfo{author}{Wellmann, J.F.}, \bibinfo{author}{Liang, R.Y.},
  \bibinfo{year}{2019}.
\newblock \bibinfo{title}{A bayesian unsupervised learning approach for
  identifying soil stratification using cone penetration data}.
\newblock \bibinfo{journal}{Canadian Geotechnical Journal}
  \bibinfo{volume}{56}, \bibinfo{pages}{1184--1205}.
\bibitem[{Wang et~al.(2018)Wang, Wang, Liang, Zhu and Di}]{Wang2018}
\bibinfo{author}{Wang, X.}, \bibinfo{author}{Wang, H.}, \bibinfo{author}{Liang,
  R.Y.}, \bibinfo{author}{Zhu, H.}, \bibinfo{author}{Di, H.},
  \bibinfo{year}{2018}.
\newblock \bibinfo{title}{A hidden markov random field model based approach for
  probabilistic site characterization using multiple cone penetration test
  data}.
\newblock \bibinfo{journal}{Structural Safety} \bibinfo{volume}{70},
  \bibinfo{pages}{128--138}.
\newblock \DOIprefix\doi{10.1016/j.strusafe.2017.10.011}.
\bibitem[{Webster(2000)}]{Webster2000}
\bibinfo{author}{Webster, R.}, \bibinfo{year}{2000}.
\newblock \bibinfo{title}{Is soil variation random?}
\newblock \bibinfo{journal}{Geoderma} \bibinfo{volume}{97},
  \bibinfo{pages}{149--163}.
\newblock \DOIprefix\doi{https://doi.org/10.1016/S0016-7061(00)00036-7}.
\bibitem[{Xie et~al.(2023)Xie, Merino, Moretti, Pauwels, Chang and
  Parlikad}]{xie2023digital}
\bibinfo{author}{Xie, X.}, \bibinfo{author}{Merino, J.},
  \bibinfo{author}{Moretti, N.}, \bibinfo{author}{Pauwels, P.},
  \bibinfo{author}{Chang, J.Y.}, \bibinfo{author}{Parlikad, A.},
  \bibinfo{year}{2023}.
\newblock \bibinfo{title}{Digital twin enabled fault detection and diagnosis
  process for building hvac systems}.
\newblock \bibinfo{journal}{Automation in Construction} \bibinfo{volume}{146},
  \bibinfo{pages}{104695}.
\bibitem[{Yoshida et~al.(2021)Yoshida, Tomizawa and
  Otake}]{yoshida2021estimation}
\bibinfo{author}{Yoshida, I.}, \bibinfo{author}{Tomizawa, Y.},
  \bibinfo{author}{Otake, Y.}, \bibinfo{year}{2021}.
\newblock \bibinfo{title}{Estimation of trend and random components of
  conditional random field using gaussian process regression}.
\newblock \bibinfo{journal}{Computers and Geotechnics} \bibinfo{volume}{136},
  \bibinfo{pages}{104179}.
\bibitem[{Zhang et~al.(2021)Zhang, Li, Li, Liu, Chen and
  Ding}]{zhang2021application}
\bibinfo{author}{Zhang, W.}, \bibinfo{author}{Li, H.}, \bibinfo{author}{Li,
  Y.}, \bibinfo{author}{Liu, H.}, \bibinfo{author}{Chen, Y.},
  \bibinfo{author}{Ding, X.}, \bibinfo{year}{2021}.
\newblock \bibinfo{title}{Application of deep learning algorithms in
  geotechnical engineering: a short critical review}.
\newblock \bibinfo{journal}{Artificial Intelligence Review} ,
  \bibinfo{pages}{1--41}.
\bibitem[{Zhang et~al.(2018)Zhang, Zhong, Wu, Guan, Yue and Wu}]{Zhang2018}
\bibinfo{author}{Zhang, Y.}, \bibinfo{author}{Zhong, D.}, \bibinfo{author}{Wu,
  B.}, \bibinfo{author}{Guan, T.}, \bibinfo{author}{Yue, P.},
  \bibinfo{author}{Wu, H.}, \bibinfo{year}{2018}.
\newblock \bibinfo{title}{3d parametric modeling of complex geological
  structures for geotechnical engineering of dam foundation based on
  t-splines}.
\newblock \bibinfo{journal}{Computer-Aided Civil and Infrastructure
  Engineering} \bibinfo{volume}{33}, \bibinfo{pages}{545--570}.
\newblock \DOIprefix\doi{10.1111/MICE.12343}.
\bibitem[{Zhou et~al.(2013)Zhou, Ding and Chen}]{Zhou2013}
\bibinfo{author}{Zhou, Y.}, \bibinfo{author}{Ding, L.Y.},
  \bibinfo{author}{Chen, L.J.}, \bibinfo{year}{2013}.
\newblock \bibinfo{title}{Application of 4d visualization technology for safety
  management in metro construction}.
\newblock \bibinfo{journal}{Automation in Construction} \bibinfo{volume}{34},
  \bibinfo{pages}{25--36}.

\end{thebibliography}

\end{document}